\documentclass[preprint,aps,preprintnumbers, showpacs, nofootinbib,superscriptaddress]{revtex4-1}
  \usepackage{latexsym,bm,amsmath,amssymb,amsfonts,color}
\newcommand{\beq}{\begin{eqnarray}}
\newcommand{\eeq}{\end{eqnarray}}

\newcommand{\Slash}[1]{{\ooalign{\hfil/\hfil\crcr$#1$}}}

\usepackage{graphicx}
\usepackage{amsmath,amsthm,amssymb}
\usepackage{color}

\begin{document}

\title{Twist-four gravitational form factor at NNLO QCD\\ from trace anomaly constraints}

\author{Kazuhiro Tanaka}
\email{kztanaka@juntendo.ac.jp}
\affiliation{Department of Physics, Juntendo University, Inzai, Chiba 270-1695, Japan}

\date{\today}

\begin{abstract}
It is known that the trace anomaly in the QCD energy-momentum tensor $T^{\mu \nu}$ can be attributed to the anomalies for each of the gauge-invariant quark part and gluon part of $T^{\mu \nu}$,
and their explicit three-loop formulas have been derived in the $\overline{\rm MS}$
scheme in the dimensional regularization. 
The
matrix elements of this quark/gluon decomposition of the QCD trace anomaly allow us to
derive the QCD constraints on the hadron's gravitational form factors, in particular, on the twist-four gravitational form factor,
$\bar{C}_{q,g}$.
Using the three-loop quark/gluon trace anomaly formulas, we
calculate the forward (zero momentum transfer) value of the twist-four gravitational form factor $\bar{C}_{q,g}$ at the next-to-next-to-leading-order (NNLO) accuracy.
We present quantitative results
for nucleon as well as for pion, 
leading to a model-independent
determination of the forward value of $\bar{C}_{q,g}$.
We find quite different pattern in the obtained results
between the nucleon and the pion.
In particular, for the nucleon, the present information from experiment and lattice QCD on the nonperturbative matrix elements 
arising in our NNLO formula allows us
to obtain a prediction of the forward value of $\bar{C}_{q,g}$ at the accuracy of a few percent
level.
\end{abstract}
\maketitle

\section{Introduction} 
\label{sec1}

The QCD energy-momentum tensor $T^{\mu\nu}$ 
is known to receive the trace anomaly~\cite{Nielsen:1977sy,Adler:1976zt,Collins:1976yq}, 
as 
\begin{equation}
T^\mu_\mu=\eta_{\mu\nu} T^{\mu\nu} =\frac{\beta(g)}{2g}F^2
+ \left(1+\gamma_m(g)\right)m\bar{\psi}\psi\ ,  
\label{111}
\end{equation}
representing the broken scale invariance due to the quantum loop effects,
with the beta-function $\beta(g)$ for the QCD coupling constant $g$ 
and the anomalous dimension $\gamma_m(g)$ for the quark mass $m$. 
Here, $\eta_{\mu\nu}$ is the 
metric tensor,  $\eta_{\mu\nu}={\rm diag}\left( 1,\ -1,\ -1,\ -1\right)$ in four spacetime dimensions,
and 
$F^2$ ($=F_a^{\mu\nu}{F_a}_{\mu\nu}$) and $m\bar{\psi}\psi$ ($=m_u \bar u u+m_d \bar d d +\ldots$) denote the renormalized composite operators dependent on a renormalization scale.
The 
symmetric QCD energy-momentum tensor is given by~\cite{Ji:1996ek} 
(see also \cite{Treiman:1986ep,Braun:2003rp})
\begin{eqnarray}
T^{\mu\nu}
=  T^{\mu\nu}_q+ T^{\mu\nu}_g 
\label{tqg}
\end{eqnarray}
where the operators,
\begin{equation}
T^{\mu\nu}_q= i\bar{\psi}\gamma^{(\mu}\overleftrightarrow{D}^{\nu)}\psi\ ,
\;\;\;\;\;\;\;\;\;\;\;\;
T_g^{\mu\nu}= -F^{\mu\lambda}F^\nu_{\ \lambda} + \frac{\eta^{\mu\nu}}{4}F^2\ ,
\label{tg}
\end{equation}
with  
$D^\mu=\partial^\mu+ig A^\mu$, $\overleftrightarrow{D}^\mu \equiv \frac{\overrightarrow{D}^\mu -\overleftarrow{D}^\mu}{2}$ and $R^{(\mu}S^{\nu)}\equiv \frac{R^\mu S^\nu+R^\nu S^\mu}{2}$,
are the gauge-invariant quark part and gluon part;
we have neglected the  gauge-variant terms in the RHS of (\ref{tqg}), i.e., the ghost term and the gauge fixing term, as they do not affect our final results.
Classically, we have, 
${\eta_{\mu \nu }}T_q^{\mu \nu } = m\bar \psi\psi$ and ${\eta_{\mu \nu }}T_g^{\mu \nu } = 0$, up to the terms that vanish by the equations of motion (EOM), 
but (\ref{111}) does not coincide with the quantum corrections to the $m\bar \psi\psi$ operator,
reflecting that renormalizing the quantum loops and taking the trace do not commute.
We note that the total tensor $T^{\mu\nu}$ of (\ref{tqg}) is not renormalized;  it is a finite, scale-independent operator, because of
the energy-momentum conservation, 
\begin{equation}
\partial_\nu T^{\mu\nu}=0\ ,
\label{divless}
\end{equation}
while $T_q^{\mu\nu}$ and $T_g^{\mu\nu}$ are not conserved separately 
and thus each of  $T_q^{\mu\nu}$ and $T_g^{\mu\nu}$ is subject to regularization and renormalization. 
This fact suggests that each of $T_q^{\mu\nu}$ and $T_g^{\mu\nu}$ should
receive
a definite amount of anomalous trace contribution, such that their sum reproduces~(\ref{111}).
The corresponding trace anomaly for each quark/gluon part
is derived up to two-loop order in \cite{Hatta:2018sqd}. The extension to the three-loop order
is worked out in \cite{Tanaka:2018nae}, 
demonstrating that the logic to determine the quark/gluon decomposition of the trace anomaly holds to all orders in perturbation theory.
In the MS-like (MS,  $\overline{\rm MS}$) schemes in the dimensional regularization,  we obtain
\begin{eqnarray}
&&\eta_{\mu \nu }T_q^{\mu \nu }  =  m\bar \psi\psi + \frac{{{\alpha _s}}}{{4\pi }}\left( {\frac{{{n_f}}}{3}{F^2} + \frac{{4{C_F}}}{3}m\bar \psi\psi} 
\right)+\cdots\ ,  \nonumber\\
&&\eta_{\mu \nu }T_g^{\mu \nu }  = \frac{{{\alpha _s}}}{{4\pi }}\left( { - \frac{{11{C_A}}}{6}{F^2} + \frac{{14{C_F}}}{3}m\bar \psi\psi} \right)+\cdots\ ,
\label{ano}
\end{eqnarray}
for $n_f$ flavor and $N_c$ color with $C_F=(N_c^2-1)/(2N_c)$ and $C_A=N_c$;
here $\alpha_s=g^2/(4\pi)$, and the ellipses stand for the two-loop (${\cal O}(\alpha_s^2)$)
as well as three-loop  (${\cal O}(\alpha_s^3)$) corrections, whose explicit formulas are presented in \cite{Hatta:2018sqd,Tanaka:2018nae}.
The sum of the two formulas of (\ref{ano}) coincides with (\ref{111})
at every order in $\alpha_s$.
For a recent extention to the four-loop order, see \cite{Ahmed:2022adh}.

Each formula of (\ref{ano}) is separately 
renormalization group (RG)-invariant up to the one-loop terms that are explicitly shown above,
but $\eta_{\mu\nu}T_q^{\mu\nu}$ as well as $\eta_{\mu\nu}T_g^{\mu\nu}$
receives the RG scale dependence beyond the one-loop order, still 
the total anomaly (\ref{111})
is scale-independent.
Although intuitive interpretation of the separate anomalies (\ref{ano}) for quark/gluon parts, as well as their correspondence to the quark/gluon degrees of freedom participating in the quantum loops, is not straightforward beyond one-loop order,
the formulas (\ref{ano}) could be useful beyond being a purely formal decomposition.
Indeed, the separate anomalies (\ref{ano}), as well as their 
RG properties, allow us to constrain the twist-four gravitational form factor $\bar{C}_{q,g}$~\cite{Hatta:2018sqd},
where $\bar{C}_{q}$ ($\bar{C}_{g}$)
arises as one of the gravitational form factors~\cite{Ji:1996ek,Kumano:2017lhr,Polyakov:2018zvc,Tanaka:2018wea}
to parametrize the hadron matirx element of each of quark and gluon parts of the QCD energy-momentum tensor, 
$\langle p'|T^{\mu\nu}_{q,g}|p\rangle$.
In particular, it has been demonstrated~\cite{Hatta:2018sqd} that the solution of the corresponding two-loop RG equations provides a 
model-independent
determination of the forward ($p'\to p$) value of $\bar{C}_{q,g}$, at the accuracy of $\sim$ ten percent level.
Such quantitative constraint could have impact on the developments to 
describe the shape deep inside the hadrons reflecting dynamics of quarks and gluons,
such as the pressure distributions inside the hadrons~\cite{Polyakov:2018guq,Polyakov:2018zvc,Teryaev:2016edw};
indeed, the recent results of the pressure distributions~\cite{Burkert:2018bqq}  and the shear force distributions~\cite{Burkert:2021ith} inside the nucleon
are based on the determination of the
quark part of the 
gravitational form factors from
the behaviors 
of the generalized parton distributions (GPDs)~\cite{Polyakov:2002yz,Diehl:2003ny,Belitsky:2005qn},
which are 
obtained by experiments like deeply virtual Compton scattering (DVCS)~\cite{Mueller:1998fv,Ji:1996nm,Radyushkin:1996nd,Goeke:2001tz,Diehl:2003ny,Belitsky:2005qn}, 
deeply virtual meson production~\cite{Collins:1996fb,Goloskokov:2009ia}, meson-induced Drell-Yan 
production~\cite{Berger:2001zn,Goloskokov:2015zsa,Sawada:2016mao}, etc.
As another phenomenological implication, 
the cross section of the near-threshold photoproduction of $J/\psi$ in $ep$ scattering, proposed to be measured at the Jefferson Laboratory~\cite{Joosten:2018gyo}, 
is sensitive to the $F^2$ part of the trace anomaly (\ref{111})~\cite{Kharzeev:1998bz},
which can be conveniently handled~\cite{Hatta:2018ina} through the $p'\to p$ behavior of 
the gravitational form factors that parametrize $\langle p'|\eta_{\mu\nu}T_{g}^{\mu\nu}|p\rangle$.  
The separate anomalies (\ref{ano}) should also provide a new insight
on understanding the origin of the nucleon mass~\cite{Ji:1994av,Lorce:2017xzd,Lorce:2018egm,Metz:2020vxd,Lorce:2021xku,Ji:2021qgo,Liu:2021gco} to be explored in the future Electron-Ion Collider.

In this paper we extend the two-loop 
calculation of \cite{Hatta:2018sqd} for constraining the twist-four gravitational form factor $\bar{C}_{q,g}$ to the calculation at the next higher-order accuracy 
by using the three-loop formulas for the separate anomalies, (\ref{ano}).
We
calculate the forward value of the twist-four gravitational form factor $\bar{C}_{q,g}$ at the next-to-next-to-leading-order (NNLO) accuracy, which allows us to present quantitative results
for nucleon as well as pion; in particular, for nucleon, we determine the forward value of $\bar{C}_{q,g}$
at the level of accuracy $\sim$ a few percent.

The rest of the paper is organized as follows. We sketch  
all-orders renormalization-mixing structure relevant for the quark and gluon energy-momentum tensors
in the MS-like schemes in Sec~\ref{sec20}, and explain, as its direct consequence,  
the quark/gluon decomposition of the QCD trace anomaly. 
Implications of this result to constrain 
the gravitational form factors are discussed in Sec.~\ref{sec4}.
In particular, in Sec.~\ref{secnucleon}, we present a model-independent
determination of the forward value of the twist-four gravitational form factor $\bar{C}_{q,g}$, at the accuracy of $\sim$ a few percent
level. These results for a nucleon are extended to the case of the spin-0 hadrons like a pion in Sec.~\ref{pionsec}.
Sec.~\ref{conc} is reserved for conclusions.

\section{Renormalization structure of energy-momentum tensor and 
the separate quark and gluon trace anomalies}  
\label{sec20}

In this section we sketch how the formulas (\ref{ano}) 
are obtained. 
First of all, the renormalization of $T_q^{\mu\nu}$, $T_g^{\mu\nu}$
of (\ref{tg}) 
is not straightforward, because it  does not obey a simple multiplicative form: 
$T_q^{\mu\nu}$, $T_g^{\mu\nu}$ are composed of 
the twist-two (traceless part) and twist-four (trace part) operators,
and
the renormalization mixing between the quark part and gluon part also arises.
To treat them, we define a basis of independent gauge-invariant operators up to twist four,
\begin{eqnarray}
&&O_g=-F^{\mu\lambda}F^{\nu}_{\ \lambda}\ , \;\;\;\;\;\;\;\;\;\;\;\;O_q=i\bar{\psi}\gamma^{(\mu}\overleftrightarrow{D}^{\nu)} \psi\ ,  
\label{ops}\\
&& O_{g(4)}= \eta^{\mu\nu}F^2\ ,\;\;\;\;\;\;\;\;\;\;\;\;
O_{q(4)}=\eta^{\mu\nu}m\bar{\psi}\psi\ ,  
\end{eqnarray}
and the corresponding bare operators, $O_k^B$ with $k=g, q, g(4)$, and $q(4)$.
The renormalization constants are introduced as
\begin{eqnarray}
&&O_g=Z_T O_g^B + Z_MO_{g(4)}^B + Z_L O_q^B + Z_S O_{q(4)}^B\ , 
\label{o3ren0}\\
&&O_q=Z_\psi O_q^B +Z_KO_{q(4)}^B+ Z_Q O_g^B +Z_B O_{g(4)}^B\ ,  \label{o3ren}
\\
&& O_{g(4)}=Z_F O_{g(4)}^B+Z_C O_{q(4)}^B\ , 
\label{o4ren0} \\
&&O_{q(4)}= O_{q(4)}^B\ , \label{o4ren} 
\end{eqnarray}
where, for simplicity, the mixing with the EOM operators as well as the BRST-exact operators is not shown, as their physical matrix elements vanish and they do not affect our final result~\cite{Kodaira:1998jn}.
Here, $O_g$, as well as $O_q$, is a mixture of the twist-two and -four operators,
and the corresponding twist-four components receive the contributions of the twist-four operators $O_{g(4)}$ and $O_{q(4)}$.
The latter two formulas (\ref{o4ren0}) and (\ref{o4ren}) reflect, respectively, that the twist-four operator $O_{g(4)}$ mixes with itself and another twist-four operator $O_{q(4)}$, and that $O_{q(4)}$ is renormalization group (RG)-invariant (see \cite{Tarrach:1981bi,Hatta:2018sqd,Tanaka:2018nae}).

Subtracting the traces from both sides of the equations~(\ref{o3ren0}) and (\ref{o3ren}), 
$O_k$ and $O_k^B$ with $k=g, q$ are replaced by the corresponding twist-two parts, $O_{k(2)}$ and $O^B_{k(2)}$, respectively,
such that the twist-four contributions drop out:
\begin{eqnarray}
 O_{g(2)}&=&Z_T  O_{g(2)}^B+
 Z_L  O_{q(2)}^B   \ ,  
\nonumber
\\
O_{q(2)}&=&Z_\psi O_{q(2)}^B + Z_Q  O_{g(2)}^B \ . \label{o1rentl}
\end{eqnarray}
Here, the resultant equations are controlled by 
the renormalization constants $Z_T, Z_L,  Z_\psi$ and $Z_Q$, and 
should represent the flavor-singlet mixing of the twist-two, spin-two operators.
Thus, those four constants $Z_T, Z_L,  Z_\psi$ and $Z_Q$ can be determined by the second moments of the Dokshitzer-Gribov-Lipatov-Altarelli-Parisi (DGLAP) splitting functions which are known up to the three-loop
accuracy~\cite{Vogt:2004mw}.~\footnote{See \cite{Moch:2021qrk} for recent four loop results for the low moments.}

For the renormalization mixing (\ref{o4ren0}) at twist four,
the Feynman diagram calculation of $Z_F$ and $Z_C$ is available to the two-loop order~\cite{Tarrach:1981bi}.
Moreover, it is shown \cite{Tanaka:2018nae} that the constraints imposed by the RG invariance of  (\ref{111})
allow to determine the power series in $\alpha_s$ for $Z_F$ as well as $Z_C$ in the MS-like schemes,
completely from the perturbative expansions of $\beta(g)$ and 
$\gamma_m(g)$,  which are now known to five-loop order~\cite{Baikov:2016tgj,Luthe:2016ima,Herzog:2017ohr,
Baikov:2014qja,Luthe:2017ttc,Luthe:2016xec} in the literature. 

Therefore, six renormalization constants $Z_T, Z_L,  Z_\psi, Z_Q, Z_F$ and $Z_C$
among ten constants arising in (\ref{o3ren0})-(\ref{o4ren}) are
available to a certain accuracy beyond two-loop order in the MS-like schemes, and they 
take the form,  
\begin{equation}
Z_X=\left(\delta_{X,T}+\delta_{X,\psi}+\delta_{X,F}\right)+\frac{a_X}{\epsilon}+\frac{b_X}{\epsilon^2}+\frac{c_X}{\epsilon^3}
+\cdots
\ ,
\label{zx}
\end{equation}
in the $d=4-2\epsilon$ spacetime dimensions with $X=T,L, \psi , Q,F$, and $C$; here, $a_X, b_X, c_X, \ldots,$ are the constants given as the power series in $\alpha_s$, and
$\delta_{X,X'}$ denotes the Kronecker 
symbol.
However, $Z_M$, $Z_S$, $Z_K$ and $Z_B$ still remain unknown.
It is shown \cite{Tanaka:2018nae} that these four renormalization constants can be determined to the accuracy same as the renormalization constants (\ref{zx}),
by invoking that they should also obey the form (\ref{zx}) with $X =M, S, K, B$, and that the RHS of the formulas~(\ref{o3ren0}), (\ref{o3ren}) are, in total, UV-finite. Thus, all the renormalization constants in  (\ref{o3ren0})-(\ref{o4ren}) are determined
up to the three-loop accuracy.
The trace part of each of the renormalized quark part and gluon part (\ref{tg}), 
$\eta_{\mu\nu}T^{\mu\nu}_{q,g}$,
is of twist four and thus is expressed as a superposition
in terms 
of the independent twist-four renormalized operators,
$m\bar{\psi}\psi$ and $F^2$.
The corresponding formulas can be derived calculating the trace part of (\ref{o3ren}) and (\ref{o3ren0}), 
and then reexpressing the results with $m\bar{\psi}\psi$ and $F^2$
by the use of (\ref{o4ren0}), (\ref{o4ren}). 
Expressing the resulting formulas as
\begin{eqnarray}
&&\eta_{\mu\nu}T^{\mu\nu}_q=x_q(\alpha_s)F^2
+ \left(1+y_q(\alpha_s)\right)m\bar{\psi}\psi\ ,
\label{tgrenaq}\\
&&\eta_{\mu\nu}T_g^{\mu\nu}= x_g(\alpha_s)F^2
+y_g(\alpha_s)m\bar{\psi}\psi\ ,
\label{tgrenag}
\end{eqnarray} 
the coefficients $x_{q}(\alpha_s)$, $y_{q}(\alpha_s)$, $x_{g}(\alpha_s)$, and $y_{g}(\alpha_s)$ are 
completely determined by the renormalization constants in (\ref{o3ren0})-(\ref{o4ren}),
such that those coefficients are the finite quantities of order $\alpha_s$ and higher.
Furthermore,
it has been shown~\cite{Tanaka:2018nae} that the relations,
\begin{eqnarray}
&&x_q(\alpha_s)+x_g(\alpha_s)=\frac{\beta(g)}{2g}\ , \nonumber\\
&&y_q(\alpha_s)+y_g(\alpha_s)=\gamma_m(g)\ ,
\label{xy}
\end{eqnarray}
hold to all orders in $\alpha_s$,
where 
\begin{eqnarray}
&&\beta (g) = \frac{dg}{d\ln \mu}=\sqrt{\frac{\pi}{\alpha_s}}\ \frac{d\alpha_s}{d\ln \mu}
=  -\sqrt{4\pi\alpha_s}\sum_{n=0}^\infty\beta_n \left(\frac{\alpha_s}{4\pi}\right)^{n+1}\ ,
\label{betaexact}\\
&&\gamma_m(g)= -\frac{1}{m} \frac{\partial m(\mu)}{\partial \ln \mu}
=\sum_{n=0}^\infty\gamma_{mn}\left(\frac{\alpha_s}{4\pi}\right)^{n+1}
\ ,
\label{betagR}
\end{eqnarray}
so that the relations (\ref{xy}) guarantee that the sum of (\ref{tgrenaq}) and (\ref{tgrenag})
reproduces the QCD trace anomaly~(\ref{111}).
We note that
the sum of the two equations (\ref{tgrenaq}) and (\ref{tgrenag}) is thus
RG-invariant; but,
each of them exhibits the dependence on the renormalization scale $\mu$ in the MS-like schemes, i.e., 
\begin{equation}
T^\lambda_\lambda=\left. \eta_{\lambda\nu}T^{\lambda\nu}_g\right|_\mu+
\left. \eta_{\lambda\nu}T^{\lambda\nu}_q\right|_\mu\ ,
\label{separatemu}
\end{equation}
due to the contributions of order $\alpha_s^2$ and higher (see the discussion in Sec.~\ref{sec1}).

The results (\ref{tgrenaq}), (\ref{tgrenag}) allow us to derive the three-loop formulas for (\ref{ano});
here, the explicit form of (\ref{tgrenaq}), (\ref{tgrenag}) is given, in the MS-like schemes,
up to the three-loop order in Eqs.~(88), (87) of \cite{Tanaka:2018nae},
and the corresponding three-loop formulas of 
$x_{q}(\alpha_s)$, $y_{q}(\alpha_s)$, $x_{g}(\alpha_s)$, and $y_{g}(\alpha_s)$ in the MS-like schemes
are given as $x_3$, $y_3$, $x_1$, and $y_1$ in Eqs.~(83)-(86) in \cite{Tanaka:2018nae}.
Therefore, through the renormalization, each of the quark part $T_q^{\mu\nu}$ and the gluon part $T_g^{\mu\nu}$ of the 
energy-momentum tensor 
receives a definite amount of anomalous trace contribution as in (\ref{tgrenaq}),  (\ref{tgrenag}), such that their sum reproduces (\ref{111}).

\section{Anomaly constraints on the nucleon’s twist-four gravitational form factor}
\label{sec4}

The nucleon matrix element of  each term in (\ref{tqg}), using the nucleon states $|N(p)\rangle$ and  $|N(p')\rangle$
with the 4-momenta $p$ and $p'$, respectively,
is parameterized as 
\begin{eqnarray}
 \langle N(p')| T_{q,g}^{\mu\nu}|N(p)\rangle &=& \bar{u}(p')\Bigl[A_{q,g} (t)\gamma^{(\mu}\bar P^{\nu)}
 +B_{q,g}(t)\frac{\bar P^{(\mu}i\sigma^{\nu)\alpha}\Delta_\alpha}{2M}
 \nonumber\\
 &+& D_{q,g}(t)\frac{\Delta^\mu\Delta^\nu -\eta^{\mu\nu}t}{4M} + \bar{C}_{q,g}(t)M\eta^{\mu\nu}\Bigr] u(p)\ ,
\label{para}
\end{eqnarray}
in terms of the gravitational form factors $A_{q,g} (t), B_{q,g}(t), D_{q,g}(t)$, and $\bar{C}_{q,g}(t)$~\cite{Polyakov:2018zvc,Tanaka:2018wea},
where $\bar{P}^\mu\equiv \frac{p^\mu+p'^\mu}{2}$ is the average of the initial and final momenta, $\Delta^\mu=p'^\mu-p^\mu$ is the momentum transfer, $t=\Delta^2$, 
and $M$ and $u(p)$ are the nucleon mass and spinor, so that
$\bar{P}^2= M^2 -t/4$.
$A_{q,g} (t)$ and $B_{q,g}(t)$ are familiar twist-two form factors;
$A_{q,g} (t)$ obey the forward ($t \to 0$) sum rule,  
\begin{equation}
A_q(0)+A_g(0) =1\ ,
\label{aqag1}
\end{equation}
representing a sharing of the total momentum by the quarks/gluons,
as a consequence of the forward matrix element of the energy-momentum tensor~(\ref{tqg}) being
normalized by 
\begin{equation}
\langle N(p)|T^{\mu\nu}|N(p)\rangle =2p^\mu p^\nu\ ,
\label{normalize} 
\end{equation}
with $\langle N(p')|N(p)\rangle =2p^0 (2\pi)^3 \delta^{(3)}(p'-p)$ being assumed,
and, similarly, $B_{q,g}(t)$ obey the forward sum rule, $[A_q(0)+B_q(0) +A_g(0) +B_g(0)]/2=1/2$, 
representing a sharing of the total angular momentum by the quarks/gluons, as a consequence of the nucleon spin being $1/2$.

$D_{q,g}(t)$, $\bar{C}_{q,g}(t)$ of (\ref{para}) have also received considerable attention recently~\cite{Burkert:2018bqq,Polyakov:2018zvc,Tanaka:2018wea,Kumano:2017lhr,Polyakov:2018guq,Teryaev:2016edw,Burkert:2021ith,Hatta:2018ina},
and their theoretical estimates are performed~\cite{Ji:1997gm,Goeke:2007fp,Pasquini:2014vua,Polyakov:2018exb,Lorce:2018egm,Shanahan:2018nnv,Shanahan:2018pib,Anikin:2019kwi,Azizi:2019ytx,Fujita:2022jus}:
$D_{q,g}(t)$ are related to 
the so-called D term, $D\equiv D_q(0)+D_g(0)$~\cite{Polyakov:2018zvc}.
For $\bar{C}_{q,g}(t)$, 
exact manipulations for the divergence of  (\ref{tg}) yield the operator identities~\cite{Kolesnichenko:1984dj,Braun:2004vf,Tanaka:2018wea},
\begin{equation}
\partial _\nu T_q^{\mu \nu } =  \bar \psi g{F^{\mu \nu }}{\gamma _\nu }\psi\ ,  \;\;\;\;\;\;\;\;\;\;
\partial _\nu T_g^{\mu \nu } =  - F_a^{\mu \nu }D_{ab}^\rho F_{\rho \nu }^b\ .  \label{phys}
\end{equation}
up to the terms which vanish using the equations of motion, $\left(i\Slash{D}-m\right)\psi=0$,
and the matrix elements of these identities can be expressd by $\bar{C}_{q,g}(t)$ using (\ref{para}), as~\cite{Tanaka:2018wea}
\begin{eqnarray}
&&\langle N(p')|g\bar{\psi}F^{\mu\nu}\gamma_\nu \psi |N(p)\rangle =iM \Delta^\mu \bar{C}_q (t)\bar{u}(p')u(p)\ , \label{mat1} \\
&&-\langle N(p')|    F_a^{\mu \nu }D_{ab}^\rho F_{\rho \nu }^b |N(p)\rangle =iM \Delta^\mu \bar{C}_g (t)\bar{u}(p')u(p)\ , \label{mat2}
\end{eqnarray}
showing that $\bar{C}_{q,g}(t)$ represent the multiparton correlation of twist four.
The identities of (\ref{phys}) are compatible with the condition (\ref{divless}),
using the equations of motion for the gluon fields, $D_\alpha F^{\alpha\nu}=g\bar{\psi}\gamma^\nu \psi$,
and the fact that the equations of motions are preserved under renormalization; therefore, we have
\begin{equation}
\bar{C}_q(t)+\bar{C}_g(t) =0\ ,
\label{cqcg0}
\end{equation}
for all values of $t$.
We note that $\bar{C}_{q,g}(t)$ are relevant to
the force distribution inside the nucleon~\cite{Polyakov:2018zvc,Burkert:2021ith}
and the nucleon's 
transverse spin sum rule~\cite{Hatta:2012jm}.

The formula (\ref{separatemu}) indicates that the gravitational form factors $A_{q,g} (t), B_{q,g}(t), D_{q,g}(t)$, and $\bar{C}_{q,g}(t)$ in (\ref{para}) depend on the renormalization scale $\mu$;
this renormalization scale dependence 
can be determined from 
the renormalization-mixing structures in (\ref{o3ren0})-(\ref{o4ren}).
The corresponding renormalization group (RG) equations of  $\bar{C}_{q,g}$ and $A_{q,g}$
and the scale evolutions implied by their solutions
are discussed at the two-loop level in \cite{Hatta:2018sqd}.
In this paper we discuss the evolutions of $\bar{C}_{q,g}$ as well as of $A_{q,g}$ at the three-loop level,
as a result of the three-loop formulas for (\ref{o3ren0})-(\ref{o4ren}), (\ref{tgrenaq}),  (\ref{tgrenag}) derived in \cite{Tanaka:2018nae}.
In the following, we treat the form factors,
\begin{eqnarray}
&&\bar{C}_{q,g}(\mu) \equiv \bar{C}_{q,g}(t=0, \mu)\ , \nonumber\\
&&A_{q,g}(\mu)\equiv A_{q,g}\left(t=0, \mu\right)\ ,
\label{camu}
\end{eqnarray}
in the forward limit $t =0$, 
denoting the renormalization scale $\mu$ dependence explicitly, and 
derive their three-loop evolutions
taking into account the above constraints~(\ref{cqcg0}), (\ref{aqag1}).

Substituting (\ref{para}) for the forward matrix element of the relations,
\begin{equation}
O_{k(2)}=T_k^{\mu \nu}-{\rm traces} \;\;\;\; \;\;\;\; \;\;\;\; ( k=q, g)\ ,
\label{trsubt}
\end{equation} 
between the operators (\ref{tg}) and  the twist-2 parts of (\ref{ops}), it is straightforward to see that
the form factor $A_{q,g}(\mu)$ of (\ref{camu})
obey the $\mu$-dependences implied by (\ref{o1rentl}): 
the differentiation of the relations (\ref{o1rentl}) with respect to the renormalization scale yields
the RG equations of the twist-two, spin-2 quark and gluon operators,
which coincide with the first moment of the DGLAP
evolution equations for the flavor-singlet part
of the unpolarized parton distribution functions,
\begin{equation}
  \frac{d}{d \ln \mu} \begin{pmatrix} O_{q(2)}(\mu)  \\  O_{g(2)}(\mu) \end{pmatrix}
=  - \widetilde{\bm{\gamma}}(\alpha_s)  \begin{pmatrix}  O_{q(2)}(\mu) \\  O_{g(2)}(\mu)\end{pmatrix} 
= 
 - \begin{pmatrix}\widetilde \gamma_{qq}(\alpha_s) &\widetilde \gamma_{qg}(\alpha_s) \\
\widetilde \gamma_{gq}(\alpha_s) &\widetilde \gamma_{gg}(\alpha_s) \end{pmatrix}   \begin{pmatrix}  O_{q(2)}(\mu) \\  O_{g(2)}(\mu)\end{pmatrix}  ,  \label{one}
\end{equation}
with the anomalous dimension matrix $\widetilde{\bm{\gamma}}(\alpha_s)$, as the first moment of the singlet DGLAP kernel.
The three-loop anomalous dimension matrix of (\ref{one}) for the twist-two flavor-singlet operators reads~\cite{Larin:1996wd,Vogt:2004mw}
\begin{equation}
\widetilde{\bm{\gamma}}(\alpha_s)\equiv \begin{pmatrix}\widetilde \gamma_{qq}(\alpha_s) &\widetilde \gamma_{qg}(\alpha_s) \\
\widetilde \gamma_{gq}(\alpha_s) &\widetilde \gamma_{gg}(\alpha_s) \end{pmatrix} 
= \frac{\alpha_s}{4\pi}\widetilde{\bm{\gamma}}_0
+ \left(\frac{\alpha_s}{4\pi}\right)^2 \widetilde{\bm{\gamma}}_1 +\left(\frac{\alpha_s}{4\pi}\right)^3 \widetilde{\bm{\gamma}}_2\ ,
\label{dglapk}
\end{equation}
where
\begin{equation}
\widetilde{\bm{\gamma}}_0
= \begin{pmatrix} \frac{16C_F}{3}\;\; &- \frac{4n_f}{3} \\ -\frac{16C_F}{3} \;\; & \frac{4n_f}{3} \end{pmatrix}\ ,
\label{dglap}
\end{equation}
\begin{eqnarray}
 \widetilde{\bm{\gamma}}_1&=&2\begin{pmatrix} \frac{376}{27}C_FC_A  -\frac{112}{27} C_F^2 -\frac{104}{27} n_f C_F \;\;\;\;\;\;\; & 
-\frac{74}{27}C_Fn_f -\frac{35}{27}C_An_f \\
-\frac{376}{27}C_F C_A + \frac{112}{27}C_F^2 + \frac{104}{27}C_F n_f \;\;\;\;\; \;\;& \frac{74}{27}C_F n_f + \frac{35}{27}C_A n_f  \end{pmatrix}\ ,
\end{eqnarray}  
and
\begin{eqnarray}
\widetilde{\gamma_2}_{\, qq}=-\widetilde{\gamma_2}_{\, gq}&=&-\frac{256}{3} \zeta (3) C_A C_F n_f-\frac{44}{9} C_A C_F n_f-128 \zeta (3) C_A
   C_F^2+\frac{128}{3} \zeta (3) C_A^2 C_F
\nonumber\\&&
-\frac{17056}{243} C_A
   C_F^2
+\frac{41840}{243} C_A^2 C_F+\frac{256}{3} \zeta (3) C_F^2
   n_f-\frac{14188}{243} C_F^2 n_f
\nonumber\\&&
-\frac{568}{81} C_F n_f^2+\frac{256 \zeta (3)
   C_F^3}{3}-\frac{1120 C_F^3}{243}\ ,
\nonumber\\
\widetilde{\gamma_2}_{\, qg}=-\widetilde{\gamma_2}_{\, gg}&=&
-\frac{208}{3} \zeta (3) C_A C_F n_f+\frac{278}{9} C_A C_F n_f+48 \zeta (3) C_A^2
   n_f-\frac{3589}{81} C_A^2 n_f
\nonumber\\&&
+\frac{2116}{243} C_A n_f^2+\frac{64}{3} \zeta
   (3) C_F^2 n_f-\frac{346}{243} C_F n_f^2-\frac{4310}{243} C_F^2 n_f
\ , \label{gamma2}
\end{eqnarray}
in the MS-like schemes.
Here, $\zeta(s)$ is the Riemann zeta-function with $\zeta(3)= 1.202056903\ldots$.
We note that, from the definition~(\ref{para}), we have the relation (see e.g., \cite{Polyakov:2018zvc,Tanaka:2018wea}),
\begin{equation}
A_q\left(\mu \right)= \sum_f \langle x \rangle_f(\mu)\ ,
\label{aqmu}
\end{equation}
where the sum is over the $n_f$ quark flavors $f=u,d, \ldots$, 
and
\begin{equation}
\langle x \rangle_f(\mu)=\int_0^1dx x \left(q_f(x, \mu)+q_{\bar{f}}(x, \mu)\right)
\label{xfmu}
\end{equation}
is the first moment of the quark and antiquark distributions of flavor $f$ at the scale $\mu$.
We have also the similar formulas for $A_g(\mu)$.

Explicit form of the solution for the RG equations of the type of (\ref{one})
has been 
discussed at the three- as well as four-loop accuracy
in \cite{Ellis:1993rb} (see also \cite{Furmanski:1981cw}). For the present case,
we obtain
\begin{eqnarray}
 \begin{pmatrix}  A_q(\mu)  \\ A_g(\mu) \end{pmatrix}
= \bm{E}(\mu , \mu_0)
   \begin{pmatrix} A_q(\mu_0) \\ A_g(\mu_0)\end{pmatrix} ,  \label{oneintegratea}
\end{eqnarray}
for a certain ``input'' scale $\mu_0$,
using the evolution operator $\bm{E}(\mu, \mu_0)$ that obeys
\begin{equation}
  \frac{d}{d \ln \mu} \bm{E}(\mu, \mu_0)
=  - \widetilde{\bm{\gamma}}(\alpha_s) \bm{E}(\mu, \mu_0)
\ , \;\;\;\;\;\;\;\;\;\;\;\;\;\;\;\;\;\;
\bm{E}(\mu_0, \mu_0)= \bm{1}\ ,
  \label{one2}
\end{equation}
with the three-loop anomalous dimension matrix (\ref{dglapk}).
Noting that the lowest-order mixing matrix (\ref{dglap}) satisfies
\begin{equation}
\widetilde{\bm{\gamma}}_0^2
=  \frac{16 C_F\, +\, 4n_f}{3}\widetilde{\bm{\gamma}}_0\ ,
\end{equation}
i.e.,
\begin{equation}
\widetilde{\bm{\gamma}}_0\left( \widetilde{\bm{\gamma}}_0-\frac{16 C_F\, +\, 4n_f}{3}\bm{1}
\right)=\bm{0}\ ,
\label{p0mm}
\end{equation}
we can define the projection operators,
\begin{eqnarray}
&&\bm{P}
= \frac{3}{16 C_F\, +\, 4n_f}\widetilde{\bm{\gamma}}_0\ ,
\nonumber\\
&&\bm{Q}
=\bm{1}-\bm{P}
\ ,
\label{mpm1}
\end{eqnarray}
which are associated with the eigenvalues 
($\left(16 C_F + 4n_f\right)/3$, $0$)
of the matrix $\widetilde{\bm{\gamma}}_0$
as
\begin{equation}
\widetilde{\bm{\gamma}}_0\bm{P}=\bm{P} \widetilde{\bm{\gamma}}_0=
\frac{16 C_F\, +\, 4n_f}{3}\bm{P}\ ,
\;\;\;\;\;\;\;\;\;\;\;\;\;\;\;\;\;\;\;
\widetilde{\bm{\gamma}}_0\bm{Q}=\bm{Q}\widetilde{\bm{\gamma}}_0= \bm{0}\ ,
\end{equation}
satisfying
\begin{equation}
\bm{P}^2=\bm{P}\ , \;\;\;\;\;\;\;\;\;\;\;\;\;\;\;\;\;\;\; \bm{Q}^2=\bm{Q}\ , \;\;\;\;\;\;\;\;\;\;\;\;\;\;\;\;\;\;\;\bm{P}\bm{Q}=\bm{0}\ .
\label{mpmlam}
\end{equation}
These projection operators allow us to express the LO solution for the evolution operator 
of (\ref{one2}) as
\begin{eqnarray}
\bm{E}_{\rm LO}(\mu, \mu_0)=\exp\left[\frac{\widetilde{\bm{\gamma}}_0}{2\beta_0}  \ln\left(\frac{\alpha_s(\mu)}{\alpha_s(\mu_0)}\right)\right]
=\left[\bm{Q}+\bm{P}\left(\frac{\alpha_s(\mu)}{\alpha_s(\mu_0)}\right)^{\frac{8 C_F\, +\, 2n_f}{3\beta_0}}\right]
\ ,
\label{LOsol0}
\end{eqnarray}
and we seek a solution for the full equation (\ref{one2}) as a perturbation 
about the LO 
solution, in a form~\cite{Ellis:1993rb,Furmanski:1981cw}
\begin{equation}
\bm{E}(\mu, \mu_0)=\bm{U}\left(\alpha_s(\mu)\right)\bm{E}_{\rm LO}(\mu, \mu_0)\bm{U}^{-1}\left(\alpha_s(\mu_0)\right)
\ , 
\label{seeks}
\end{equation}
with the matrices $\bm{U}\left(\alpha_s(\mu)\right)$ 
and $\bm{U}^{-1}\left(\alpha_s(\mu_0)\right)$ determined 
order-by-order, as
\begin{eqnarray}
&& \bm{U}\left(\alpha_s(\mu)\right)=\bm{1}+\frac{\alpha_s(\mu)}{4\pi}  \bm{U}_1+\left(\frac{\alpha_s(\mu)}{4\pi} \right)^2\bm{U}_2\ ,
\nonumber\\
&&\bm{U}^{-1}\left(\alpha_s(\mu_0)\right)=\bm{1}-\frac{\alpha_s(\mu_0)}{4\pi}    \bm{U}_1+\left(\frac{\alpha_s(\mu_0)}{4\pi} \right)^2\left(- \bm{U}_2+\bm{U}_1^2
\right)
\ ,
 \label{usol}
\end{eqnarray}
where
\begin{eqnarray}
\bm{U}_n
&=&\frac{1}{2n\beta_0}\left(\bm{Q}\bm{R}_n\bm{Q}+\bm{P}\bm{R}_n\bm{P}\right)
+\frac{1}{\frac{16 C_F\, +\, 4n_f}{3}+2n\beta_0}\bm{Q}\bm{R}_n\bm{P}
\nonumber\\
&&-\frac{1}{\frac{16 C_F\, +\, 4n_f}{3}-2n\beta_0}\bm{P}\bm{R}_n\bm{Q}
\ ,
\label{unsolexplicit}
\end{eqnarray}
for $n=1,2$, with
\begin{eqnarray}
&&\bm{R}_1
=\widetilde{\bm{\gamma}}_1-\frac{\beta_1}{\beta_0} \widetilde{\bm{\gamma}}_0\ , \nonumber\\
&&\bm{R}_2=
\widetilde{\bm{\gamma}}_2-\frac{\beta_1}{\beta_0}\widetilde{\bm{\gamma}}_1
-\left(\frac{\beta_2}{\beta_0}
-\frac{\beta_1^2}{\beta_0^2}\right)\widetilde{\bm{\gamma}}_0
+\bm{R}_1 \bm{U}_1
\ .
\end{eqnarray}
Thus, the evolution operator $\bm{E}(\mu, \mu_0)$ of (\ref{oneintegratea}) 
at the three-loop accuracy is expressed as
\begin{eqnarray}
\bm{E}(\mu, \mu_0)&&=
\bm{Q}+\bm{P}\left(\frac{\alpha_s(\mu)}{\alpha_s(\mu_0)}\right)^{\frac{8 C_F\, +\, 2n_f}{3\beta_0}}
\nonumber\\
&&
+\frac{\alpha_s(\mu)}{4\pi}  \bm{U}_1\bm{Q}-\frac{\alpha_s(\mu_0)}{4\pi}  \bm{Q}\bm{U}_1
+\left(\frac{\alpha_s(\mu)}{\alpha_s(\mu_0)}\right)^{\frac{8 C_F\, +\, 2n_f}{3\beta_0}}
\left[
\frac{\alpha_s(\mu)}{4\pi}    \bm{U}_1\bm{P}-\frac{\alpha_s(\mu_0)}{4\pi}  \bm{P}\bm{U}_1
\right]
\nonumber\\
&&
+\left(\frac{\alpha_s(\mu)}{4\pi} \right)^2\bm{U}_2\bm{Q}
-\frac{\alpha_s(\mu)\alpha_s(\mu_0)}{(4\pi)^2}\bm{U}_1\bm{Q} \bm{U}_1+\left(\frac{\alpha_s(\mu_0)}{4\pi} \right)^2\bm{Q}\left(- \bm{U}_2+\bm{U}_1^2
\right)
\nonumber\\
&&+\left(\frac{\alpha_s(\mu)}{\alpha_s(\mu_0)}\right)^{\frac{8 C_F\, +\, 2n_f}{3\beta_0}}
\left[
\left(\frac{\alpha_s(\mu)}{4\pi} \right)^2\bm{U}_2\bm{P}
-\frac{\alpha_s(\mu)\alpha_s(\mu_0)}{(4\pi)^2}\bm{U}_1\bm{P} \bm{U}_1
\right.
\nonumber\\
&&\left. 
+\left(\frac{\alpha_s(\mu_0)}{4\pi} \right)^2\bm{P}\left(- \bm{U}_2+\bm{U}_1^2
\right)
\right]
\ .
\label{seeksrst}
\end{eqnarray}
Here, the first and the second lines are the LO terms and the NLO terms, respectively, which are controlled at the one-loop and two-loop accuracy; the third line and the following lines denote the NNLO terms derived from the three-loop contributions.
We can confirm that $A_{q,g}(\mu)$ obeying (\ref{oneintegratea}) with (\ref{seeksrst}) satisfies
 (\ref{aqag1}), when it is satisfied at a certain scale $\mu_0$.
 Therefore,  (\ref{oneintegratea}) with (\ref{seeksrst}) reduces to
\begin{eqnarray}
&&A_q(\mu) =1- A_g(\mu) 
= \left(1, 0\right)\cdot \bm{E}(\mu , \mu_0)
  \begin{pmatrix} A_q(\mu_0) \\ 1-A_q(\mu_0)\end{pmatrix} 
  \nonumber\\
  &&
  = A_q^{\rm LO}(\mu)+A_q^{\rm NLO}(\mu)+ A_q^{\rm NNLO}(\mu)\ ,
  \label{aall}
\end{eqnarray}
where
\begin{eqnarray}
&&
A_q^{\rm LO}(\mu)
  = \frac{n_f}{4C_F+n_f} 
   +\frac{4 C_F A_q\left(\mu _0\right)+n_f \left(A_q\left(\mu _0\right)-1\right)}{4 C_F+n_f}
   \left(\frac{\alpha _s\left(\mu \right)}{\alpha _s(\mu_0 )}\right)^{\frac{ 8 C_F+2n_f}{3 \beta_0}}\ ,
\label{alo}\\
&&A_q^{\rm NLO}(\mu)=
\left(\frac{\alpha_s(\mu)}{4\pi}  \right)
\frac{4 C_F n_f \left(-59 C_A+102 C_F+26 n_f\right)}{9 \left(4
   C_F+n_f\right) \left(-3 \beta _0+8 C_F+2 n_f\right)}
\nonumber\\
&&
-\left(\frac{\alpha_s(\mu)}{\alpha_s(\mu_0)}\right)^{\frac{8 C_F\, +\, 2n_f}{3\beta_0}}
\Biggl[
\left(\frac{\alpha_s(\mu)}{4\pi} \right)   
\frac{4 C_F A_q\left(\mu
   _0\right)+n_f \left(A_q\left(\mu _0\right)-1\right)}{27 \beta _0^2
   \left(4 C_F+n_f\right)}\Bigl(
   C_F
   \left\{-376 \beta _0 C_A
  \right.
   \nonumber\\
   &&\left.
   +72 \beta _1+30 \beta _0 n_f\right\}
   -35 \beta _0 C_A
   n_f+112 \beta _0 C_F^2+18 \beta _1 n_f  \Bigr)
\nonumber\\
&&
+\left(\frac{\alpha_s(\mu_0)}{4\pi}  \right)
\frac{1}{27 \beta _0^2 \left(-3
   \beta _0+8 C_F+2 n_f\right)}\Bigl(-16 C_F^2 \left\{\left(-188 \beta _0 C_A-21 \beta
   _0^2+36 \beta _1\right) A_q\left(\mu _0\right)
\right.
\nonumber\\
&&   \left.   
   +\beta _0 n_f \left(29
   A_q\left(\mu _0\right)-14\right)\right\}-8 \beta _0 C_A C_F \left\{n_f
   \left(94-129 A_q\left(\mu _0\right)\right)+141 \beta _0 A_q\left(\mu
   _0\right)\right\}
\nonumber\\
&&   
   +6 C_F \left\{-10 \beta _0 n_f^2 \left(A_q\left(\mu
   _0\right)-1\right)+24 \beta _1 n_f \left(1-2 A_q\left(\mu
   _0\right)\right)+\beta _0^2 n_f \left(15 A_q\left(\mu _0\right)+37\right)
 \right.
\nonumber
\\
&&   \left.     
\!\!
   +36
   \beta _0 \beta _1 A_q\left(\mu _0\right)\right\}+n_f \left(35 \beta _0 C_A-18
   \beta _1\right) \left(A_q\left(\mu _0\right)-1\right) \left(2 n_f-3 \beta
   _0\right)-896 \beta _0 C_F^3 A_q\left(\mu _0\right)
\!
\Bigr)
\!
\Biggr] 
\!
,
\label{anlo}
\\
&&
A_q^{\rm NNLO}(\mu)=
-\left(\frac{\alpha_s(\mu)}{4\pi} \right)^2
\frac{C_F n_f}{81 \left(4 C_F+n_f\right) \left(4 C_F+n_f-3 \beta
   _0\right) \left(8 C_F+2 n_f-3 \beta _0\right)}
  \Bigl(
  48 \bigl\{ -1241
\nonumber\\
&&  
  +1728 \zeta (3)\bigr\} C_F^3
   +2 \left\{(57928-98496
   \zeta (3)) C_A+n_f (-26134+31104 \zeta (3))
   +81 \beta _0 (85
       \right.
\nonumber\\
&&   \left.        
   -192 \zeta
   (3))\right\} C_F^2
   +\left\{ 24 (-1951+4752 \zeta (3)) C_A^2+\left(-90720 \zeta (3)
   n_f+51268 n_f
   -9726 \beta _0
 \right.
\right.
\nonumber\\
&&   \left. 
\left.         
   +73872 \beta _0 \zeta (3)\right) C_A    
   -5508 \beta _1-3
   n_f \beta _0 (-3893+5184 \zeta (3))+2 n_f^2 (-6377
   +5184 \zeta (3))\right\} C_F
\nonumber\\
&&   
   -6
   n_f \left\{142 n_f^2-213 \beta _0 n_f+234 \beta _1\right\}+C_A^2 \left\{ 8
   (-593+3564 \zeta (3)) n_f  
   +\beta _0 (921-42768 \zeta (3))\right\}
\nonumber\\
&&   
   +C_A
   \left\{ (5458-10368 \zeta (3)) n_f^2+3 \beta _0 (-1819+5184 \zeta (3)) n_f+3186
   \beta _1\right\}
   \Bigr)
\nonumber\\
&&+\left(\frac{\alpha_s(\mu)}{\alpha_s(\mu_0)}\right)^{\frac{8 C_F\, +\, 2n_f}{3\beta_0}}
\Biggl[
\left(\frac{\alpha_s(\mu)}{4\pi} \right)^2
\frac{n_f \left(A_q\left(\mu
   _0\right)-1\right)+4 C_F A_q\left(\mu _0\right)}{2916 \left(4
   C_F+n_f\right) \beta _0^4}
   \Bigl(
   25088 \beta _0^2
   C_F^4
\nonumber\\
&&   \left.            
   +32 \beta _0 \left\{3 \left(648 \zeta (3) \beta _0^2
   -35 \beta _0^2+140 n_f
   \beta _0+336 \beta _1\right)-5264 C_A \beta _0\right\} C_F^3
\right.
\nonumber\\
&&   \left. 
   +2 \left\{3 n_f
   (-4939+7776 \zeta (3)) \beta _0^3
   +141376 C_A^2 \beta _0^2+900 n_f^2 \beta
   _0^2+8352 n_f \beta _1 \beta _0
   \right.
\right.
\nonumber\\
&&   \left. 
\left.              
   -16 C_A \left(3 [533+972 \zeta (3)] \beta
   _0^2+1900 n_f \beta _0
   +3384 \beta _1\right) \beta _0+432 \beta _1 \left(7 \beta
   _0^2+12 \beta _1\right)\right\} C_F^2
       \right.
\nonumber\\
&&   
\left.                
   +2 \left\{8 C_A^2 \left(3290 n_f+3 \beta _0
   [2615+648 \zeta (3)]\right) \beta _0^2
   -3 C_A \left(700 \beta _0 n_f^2+9 \left[3
   (161+72 \zeta (3)) \beta _0^2
       \right.
\right.\right.\right.
\nonumber\\
&&   \left. \left.\left.
\left.                
   +688 \beta _1\right] n_f+3384 \beta _0 \beta
   _1\right) \beta _0+3 \left(\left[360 \beta _0 \beta _1
   -679 \beta _0^3\right]
   n_f^2+54 \beta _1 \left[5 \beta _0^2+16 \beta _1\right] n_f
        \right.
\right.\right.
\nonumber\\
&&   \left. \left.
\left.                
   +648 \beta _0
   \left[\beta _1^2-\beta _0 \beta _2\right]\right)\right\} C_F+n_f \left\{C_A^2
   \left(2450 n_f
   +9 \beta _0 [3589-3888 \zeta (3)]\right) \beta _0^2
         \right.
\right.
\nonumber\\
&&   \left. 
\left.                 
   -6 C_A
   \left(1058 n_f \beta _0^2+315 \beta _1 \beta _0+420 n_f \beta _1\right) \beta
   _0+324 \left(2 n_f \beta _1^2
   +3 \beta _0 \left[\beta _1^2-\beta _0 \beta
   _2\right]\right)\right\} \right.
   \Bigr)
\nonumber
\end{eqnarray}
\begin{eqnarray}
&&
-\left(\frac{\alpha_s(\mu)}{4\pi}\right)\left(\frac{\alpha_s(\mu_0)}{4\pi}\right)
\frac{1}{729 \left(8 C_F+2
   n_f-3 \beta _0\right) \beta _0^4}
   \Bigl(
\left\{112 \beta _0 C_F^2+\left(-376 C_A \beta
   _0+30 n_f \beta _0
           \right.
\right.
\nonumber\\
&&   \left. 
\left.                  
   +72 \beta _1\right) C_F-35 C_A n_f \beta _0+18 n_f \beta
   _1\right\} \left\{896 \beta _0 A_q\left(\mu _0\right) C_F^3+16 \left(\left[-21
   \beta _0^2-188 C_A \beta _0
              \right.
\right.\right.
\nonumber\\
&&   \left. \left.
\left.                  
   +36 \beta _1\right] A_q\left(\mu _0\right)+n_f \beta
   _0 \left[29 A_q\left(\mu _0\right)-14\right]\right) C_F^2+8 C_A \beta _0
   \left(n_f \left[94-129 A_q\left(\mu _0\right)\right]
                 \right.
\right.
\nonumber\\
&&   \left. 
\left.                  
   +141 \beta _0 A_q\left(\mu
   _0\right)\right) C_F-6 \left(-10 \beta _0 \left[A_q\left(\mu _0\right)-1\right]
   n_f^2+24 \beta _1 \left[1-2 A_q\left(\mu _0\right)\right] n_f
                   \right.
\right.
\nonumber\\
&&   \left. 
\left.                   
   +\beta _0^2
   \left[15 A_q\left(\mu _0\right)+37\right] n_f+36 \beta _0 \beta _1 A_q\left(\mu
   _0\right)\right) C_F-n_f \left(2 n_f-3 \beta _0\right) \left(35 C_A \beta _0
                   \right.
\right.
\nonumber\\
&&   \left. 
\left.                      
   -18
   \beta _1\right) \left(A_q\left(\mu _0\right)-1\right)\right\}
   \Bigr)
+\left(\frac{\alpha_s(\mu_0)}{4\pi} \right)^2
\frac{1}{2916
   \left(4 C_F+n_f-3 \beta _0\right) \beta _0^4}
   \Bigl(
100352 \beta _0^2 A_q\left(\mu
   _0\right) C_F^5
\nonumber\\
&&   
   -128 \beta _0 \left\{\left(3 [161+648 \zeta (3)] \beta _0^2+5264
   C_A \beta _0-1008 \beta _1\right) A_q\left(\mu _0\right)-28 n_f \beta _0
   \left(22 A_q\left(\mu _0\right)-7\right)\right\} C_F^4
\nonumber\\
&&   
   +8 \left\{141376 C_A^2
   A_q\left(\mu _0\right) \beta _0^2+60 n_f^2 \left(43 A_q\left(\mu
   _0\right)-28\right) \beta _0^2-16 C_A \left(-3 [1849+972 \zeta (3)]
   A_q\left(\mu _0\right) \beta _0^2
\right.
\right.
\nonumber\\
&&   \left. 
\left.                      
   +4 n_f \left[804 A_q\left(\mu
   _0\right)-329\right] \beta _0+3384 \beta _1 A_q\left(\mu _0\right)\right) \beta
   _0+36 \left([-35+648 \zeta (3)] \beta _0^4-420 \beta _1 \beta _0^2
\right.
\right.
\nonumber
\\
&&   \left. 
\left.            
+144 \beta
   _1^2\right) A_q\left(\mu _0\right)+3 n_f \left(\left[-10368 \zeta (3)
   A_q\left(\mu _0\right)+3399 A_q\left(\mu _0\right)+2592 \zeta (3)-2212\right]
   \beta _0^3
\right.
\right.
\nonumber\\
&&   \left. 
\left.                 
   +96 \beta _1 \left[43 A_q\left(\mu _0\right)-14\right] \beta
   _0\right)\right\} C_F^3+2 \left\{32 C_A^2 \left(94 n_f \left[82 A_q\left(\mu
   _0\right)-47\right]-3 \beta _0 [7033
\right.
\right.
\nonumber\\
&&   \left. 
\left.                 
   +648 \zeta (3)] A_q\left(\mu
   _0\right)\right) \beta _0^2-4 C_A \left(100 \beta _0 \left[97 A_q\left(\mu
   _0\right)-76\right] n_f^2-3 \left[\left(5832 \zeta (3) A_q\left(\mu
   _0\right)
\right.
\right.\right.\right.
\nonumber\\
&&   \left. \left.\left.
\left.                 
   +14079 A_q\left(\mu _0\right)-3888 \zeta (3)+3844\right) \beta _0^2+48
   \beta _1 \left(94-223 A_q\left(\mu _0\right)\right)\right] n_f+36 \beta _0
   \left[\beta _0^2 (533
\right.
\right.\right.
\nonumber\\
&&   \left. \left.
\left.                    
   +972 \zeta (3))-1410 \beta _1\right] A_q\left(\mu
   _0\right)\right) \beta _0+3 \left(300 \beta _0^2 \left[A_q\left(\mu
   _0\right)-1\right] n_f^3+\left[\left(-7776 \zeta (3) A_q\left(\mu
   _0\right)
\right.
\right.\right.\right.
\nonumber\\
&&   \left. \left.\left.
\left.            
   +6755 A_q\left(\mu _0\right)+7776 \zeta (3)-7159\right) \beta _0^3+96
   \beta _1 \left(44 A_q\left(\mu _0\right)-29\right) \beta _0\right] n_f^2
\right.
\right.
\nonumber\\
&&   \left. 
\left.         
   +3
   \left[\left(7776 \zeta (3) A_q\left(\mu _0\right)-4939 A_q\left(\mu
   _0\right)+2592 \zeta (3)-2155\right) \beta _0^4-24 \beta _1 \left(145
   A_q\left(\mu _0\right)+32\right) \beta _0^2
\right.
\right.\right.
\nonumber\\
&&   \left. \left.
\left.           
   +576 \beta _1^2 \left(3 A_q\left(\mu
   _0\right)-1\right)\right] n_f+432 \beta _0 \left[7 \beta _1 \beta _0^2+6 \beta
   _2 \beta _0-18 \beta _1^2\right] A_q\left(\mu _0\right)\right)\right\} C_F^2
\nonumber\\
&&   
   +2
   \left\{2 C_A^2 \left(70 \left[223 A_q\left(\mu _0\right)-188\right] n_f^2+3
   \beta _0 \left[9072 \zeta (3) A_q\left(\mu _0\right)-34387 A_q\left(\mu
   _0\right)+2592 \zeta (3)
\right.
\right.\right.
\nonumber\\
&&   \left. \left.
\left.              
   +17040\right] n_f+36 \beta _0^2 [2615+648 \zeta (3)]
   A_q\left(\mu _0\right)\right) \beta _0^2-3 C_A \left(700 \beta _0
   \left[A_q\left(\mu _0\right)-1\right] n_f^3
\right.
\right.
\nonumber\\
&&   \left. 
\left.              
   +\left[\left(-1944 \zeta (3)
   A_q\left(\mu _0\right)-10679 A_q\left(\mu _0\right)+1944 \zeta (3)+2807\right)
   \beta _0^2+48 \beta _1 \left(164 A_q\left(\mu _0\right)
\right.
\right.\right.\right.
\nonumber\\
&&   \left. \left.\left.
\left.              
   -129\right)\right]
   n_f^2+9 \beta _0 \left[9 \left(72 \zeta (3) A_q\left(\mu _0\right)+161
   A_q\left(\mu _0\right)+312 \zeta (3)-139\right) \beta _0^2+4 \beta _1
   \left(352
\right.
\right.\right.\right.
\nonumber\\
&&   \left. \left.\left.
\left.              
   -645 A_q\left(\mu _0\right)\right)\right] n_f+10152 \beta _0^2 \beta
   _1 A_q\left(\mu _0\right)\right) \beta _0+3 \left(\beta _0 \left[679 \beta
   _0^2+360 \beta _1\right] \left[A_q\left(\mu _0\right)-1\right] n_f^3
\right.
\right.
\nonumber\\
&&   \left. 
\left.              
   -3
   \left[\left(679 A_q\left(\mu _0\right)+173\right) \beta _0^4+6 \beta _1
   \left(75 A_q\left(\mu _0\right)+29\right) \beta _0^2+144 \beta _1^2 \left(2-3
   A_q\left(\mu _0\right)\right)\right] n_f^2
\right.
\right.
\nonumber\\
&&   \left. 
\left.              
   +54 \beta _0 \left[\beta _1 \left(15
   A_q\left(\mu _0\right)+37\right) \beta _0^2+12 \beta _2 \left(2 A_q\left(\mu
   _0\right)-1\right) \beta _0+36 \beta _1^2 \left(1-2 A_q\left(\mu
   _0\right)\right)\right] n_f
\right.
\right.
\nonumber\\
&&   \left. 
\left.              
   -1944 \beta _0^2 \left[\beta _0 \beta _2-\beta
   _1^2\right] A_q\left(\mu _0\right)\right)\right\} C_F+n_f \left\{n_f-3 \beta
   _0\right\} \left\{C_A^2 \left(2450 n_f+9 \beta _0 [-3589
\right.
\right.
\nonumber\\
&&   \left. 
\left.              
   +3888 \zeta (3)]\right)
   \beta _0^2+6 C_A \left(1058 n_f \beta _0^2+315 \beta _1 \beta _0-420 n_f \beta
   _1\right) \beta _0+324 \left(2 n_f \beta _1^2
\right.
\right.
\nonumber\\
&&   \left. 
\left.             
   +3 \beta _0 \left[\beta _0 \beta
   _2-\beta _1^2\right]\right)\right\} \left\{A_q\left(\mu _0\right)-1\right\}
   \Bigr)
\Biggr]
\ ,
\label{annlo}
\end{eqnarray}
where ``$(1, 0)\cdot \ $'' in the first line of (\ref{aall}) denotes the projection onto a unit vector $(1, 0)$;
(\ref{alo}), (\ref{anlo}), and (\ref{annlo}) 
show the LO, NLO, and NNLO contributions, respectively, which are derived from the first line, the second line, and the third and the following lines of (\ref{seeksrst}).

To determine the behavior of $\bar{C}_{q,g}(\mu)$, we proceed similarly as being handled at the two-loop order 
in \cite{Hatta:2018sqd}. We present the relevant formulas in the forms that hold to all orders in $\alpha_s$.
Considering the forward ($\Delta =0$) limit of  (\ref{para}) 
at the renormalization scale $\mu$ and taking their trace part, 
we have
\begin{equation}
{\bar C}_{q,g}(\mu)
= -\frac{1}{4}A_{q,g}(\mu) +\frac{1}{8M^2}\langle N(p)|\left. \eta_{\lambda \nu}T_{q,g}^{\lambda \nu}\right|_\mu|N(p)\rangle\ ,
\label{trace}
\end{equation}
where the first term is associated with the twist-two quantities corresponding to the quark/gluon average momentum fraction
as in (\ref{aqmu}), (\ref{xfmu}); this may be interpreted as the ``twist-four target mass effects''  $-\frac{M^2}{4}A_{q,g}(\mu)$ divided by $M^2$.
The second term represents the effects due to the twist-four operators. 
Substituting the trace anomalies of (\ref{tgrenaq}), (\ref{tgrenag}), we obtain
\begin{eqnarray}
&&\bar{C}_q(\mu)= - \frac{1}{4}A_q(\mu)+ x_q(\alpha_s)\frac{\langle N(p)|F^2|N(p)\rangle}{8M^2}
+ \left(1+y_q(\alpha_s)\right)\frac{\langle N(p)|m\bar{\psi}\psi|N(p)\rangle}{8M^2}
\ ,
\label{qanqanqan}
\nonumber\\
&&\bar{C}_g(\mu)=- \frac{1}{4}A_g(\mu)+ x_g(\alpha_s)\frac{\langle N(p)|F^2|N(p)\rangle}{8M^2}
+y_g(\alpha_s)\frac{\langle N(p)|m\bar{\psi}\psi|N(p)\rangle}{8M^2}
\ . \label{gangangan} 
\end{eqnarray}
Adding these two formulas and using (\ref{aqag1}) and (\ref{xy}), we get
\begin{eqnarray}
\bar{C}_q(\mu)+\bar{C}_g(\mu) &=& - \frac{1}{4}\left(A_q(\mu)+ A_g(\mu)\right)
+\left(  x_q(\alpha_s)+x_g(\alpha_s)\right) \frac{\langle N(p)|F^2|N(p)\rangle}{8M^2}
\nonumber\\
&&+ \left(1+y_q(\alpha_s)+y_g(\alpha_s)\right)\frac{\langle N(p)|m\bar{\psi}\psi|N(p)\rangle}{8M^2}
\nonumber\\
=& -& \frac{1}{4}+\frac{1}{8M^2}\langle N(p)| \left( \frac{\beta(g)}{2g}F^2
+ \left(1+\gamma_m(g)\right)m\bar{\psi}\psi\right) |N(p)\rangle
\ , \label{ganganganadd} 
\end{eqnarray}
which shows that (\ref{cqcg0}) is satisfied when we
use the relation,
\begin{equation}
2M^2=\langle N(p)|T^\lambda_\lambda|N(p)\rangle
=\langle N(p)| \left( \frac{\beta (g)}{2g}F^2 + \left(1+\gamma_m(g)\right) m\bar{\psi} \psi \right)|N(p)\rangle\ .
\label{mass}
\end{equation}
This is nothing but the well-known
formula for the nucleon mass~\cite{Shifman:1978zn,Ji:1995sv,Tanaka:2018nae} as a consequence of  the total trace anomaly~(\ref{111})  combined with the normalization condition~(\ref{normalize}).
This fact indicates that it is important to
impose the constraint~(\ref{mass})
when evaluating each equation in (\ref{gangangan}).
Because (\ref{betaexact}) 
reads, in perturbation theory,
\begin{eqnarray}
&&\frac{2g}{\beta(g)}=\frac{-2}{\left[\beta_0 \frac{\alpha_s}{4\pi}+\beta_1 \left(\frac{\alpha_s}{4\pi}\right)^2+\beta_2 \left(\frac{\alpha_s}{4\pi}\right)^3+\cdots\right]}
\nonumber\\
&&=-\frac{2}{\beta_0} \left(\frac{4\pi}{\alpha_s}\right)\frac{1}{1+\frac{\beta_1}{\beta_0} \left(\frac{\alpha_s}{4\pi}\right)+\frac{\beta_2}{\beta_0} \left(\frac{\alpha_s}{4\pi}\right)^2+\cdots}
\ .
\label{zfzfzfh}
\end{eqnarray}
the constraint~(\ref{mass})
implies $\left\langle N(p) \right| F^2   \left| N(p)  \right\rangle\sim M^2/\alpha_s$.
We take into account the corresponding constraint exactly by eliminating 
$\left\langle N(p)  \right| F^2   \left| N(p)  \right\rangle$ in favor of $M^2$
using (\ref{mass}), 
as
\begin{equation}
\frac{\langle N(p) |F^2 |N(p) \rangle}{8M^2} =\frac{2g}{\beta (g)}\left(\frac{1}{4}- \left(1+\gamma_m(g)\right)
\frac{\langle N(p) | m\bar{\psi} \psi|N(p) \rangle}{8M^2}\right)\ ,
\label{massmass}
\end{equation}
and the substitution of this formula into (\ref{gangangan}) leads to
\begin{eqnarray}
\bar{C}_q(\mu)=&&-\bar{C}_g(\mu)
= - \frac{1}{4}A_q(\mu)+ x_q(\alpha_s)
\left\{
\frac{g}{2\beta (g)}
-\frac{2g}{\beta (g)} \left(1+\gamma_m(g)\right)
\frac{\langle N(p) | m\bar{\psi} \psi|N(p) \rangle}{8M^2}
\right\}
\nonumber\\
&&
\;\;\;\;\;\;\;\;
+ \left(1+y_q(\alpha_s)\right)\frac{\langle N(p) |m\bar{\psi}\psi|N(p) \rangle}{8M^2}
\nonumber\\
&&= \frac{1}{4}\left(-A_q(\mu)+   x_q(\alpha_s)\frac{2g}{\beta (g)}
\right)
\nonumber\\
&&
+ \left\{1+y_q(\alpha_s)-x_q(\alpha_s)\frac{2g}{\beta (g)} \left(1+\gamma_m(g)\right)\right\}\frac{\langle N(p) |m\bar{\psi}\psi|N(p) \rangle}{8M^2}
\ .
\label{qangan}
\end{eqnarray}
Combining (\ref{qangan}) with (\ref{oneintegratea}) and (\ref{seeksrst}),
we can determine the value of $\bar{C}_{q,g}(\mu)$ for arbitrary $\mu$,
to the desired accuracy.
Substituting the three-loop formulas of $x_q(\alpha_s)$ and $y_q(\alpha_s)$,
discussed in Sec.~\ref{sec20},
the mass anomalous dimension (\ref{betagR}) to the three-loop accuracy with~\cite{Chetyrkin:1997dh,Vermaseren:1997fq} 
\begin{eqnarray}
 \gamma_{m0}&=&6C_F\ ,
\label{gammam0}\\
\gamma_{m1}&=& 
3C_F^2 + \frac{97}{3}C_F C_A -\frac{10}{3}C_F n_f\ ,
\\
\gamma_{m2}&=& 
n_f \left[\left(-48 \zeta (3)-\frac{556}{27}\right) C_A C_F+(48 \zeta (3)-46)
   C_F^2\right]
\nonumber\\ &&
-\frac{129}{2} C_A C_F^2 
+\frac{11413}{54} C_A^2 C_F-\frac{70}{27}
   C_F n_f^2+129 C_F^3
\ ,
\label{gammam2}
\end{eqnarray}
and (\ref{aall}) for $A_q(\mu)$ at the three-loop accuracy, 
(\ref{qangan}) reads
\begin{eqnarray}
&&\bar{C}_q(\mu)=-\bar{C}_g(\mu)=
- \frac{1}{4} \left( \frac{n_f}{4C_F+n_f} + \frac{2n_f}{3\beta_0}\right) + \frac{1}{4}\left(\frac{2 n_f}{3 \beta _0}+1\right)\frac{\left\langle N(p) \right| m \bar{\psi }\psi   \left| N(p) \right\rangle}{2M^2}
   \nonumber
   \\
   &&\;\;\;\;\;\;\;\;\;\;\;\;\;\;\;\;\;\;\;\;\;\;\;\;\;\;
   -\frac{4 C_F A_q\left(\mu _0\right)+n_f \left(A_q\left(\mu _0\right)-1\right)}{4(4 C_F+n_f)}\left(\frac{\alpha _s\left(\mu
   \right)}{\alpha _s(\mu_0 )}\right)^{\frac{ 8 C_F+2n_f}{3 \beta_0}}
    \nonumber
    \\
   &&
\;\;\;\;\;\;\;\;\;\;\;\;\;\;\;\;
+ \frac{\alpha_s(\mu)}{4\pi}\left(
-\frac{n_f \left(34
   C_A+49 C_F\right)}{108\beta _0}+\frac{ \beta_1
   n_f}{6 \beta _0^2}
   \right.
    \nonumber
    \\ 
   &&
   \left.
 \;\;\;\;\;\;\;\;\;\;\;\;   +
 \left[\frac{n_f \left(34 C_A+157
   C_F\right)}{108\beta _0}+\frac{C_F}{3}-\frac{\beta_1 n_f}{6 \beta _0^2}\right]\frac{\left\langle N(p) \right| m \bar{\psi }\psi   \left| N(p) \right\rangle}{2M^2}
\right)-\frac{1}{4}A_q^{\rm NLO}(\mu)
\nonumber
\\
&&+\left(\frac{\alpha_s(\mu)}{4\pi}\right)^2\Biggl(
\frac{n_f^2}{\beta _0}\left[\frac{697 C_A}{1458}+\frac{169 C_F}{2916}\right]
  +n_f
   \left[
   \frac{17 \beta _1 C_A}{54\beta
   _0^2}+\frac{\beta _2}{6\beta
   _0^2}+\frac{49 \beta _1 C_F}{108\beta
   _0^2}
        \right.
    \nonumber\\
   &&
   \left.  
   +\frac{1}{\beta
   _0}
  \left\{ \left(\frac{401}{648}-\frac{26 \zeta (3)}{9}\right) C_A C_F+\left(2 \zeta
   (3)-\frac{67}{27}\right) C_A^2+\left(\frac{8 \zeta (3)}{9}-\frac{2407}{2916}\right) C_F^2\right\}
   -\frac{\beta _1^2}{6 \beta _0^3}\right]
    \nonumber\\
   &&
+
   \left[-\frac{n_f^2}{\beta _0}\left(\frac{697 C_A}{1458}+\frac{1789 C_F}{2916}\right)
   +n_f
   \left(-\frac{17\beta _1 C_A}{54\beta
   _0^2} -\frac{\beta _2}{6\beta
   _0^2}-\frac{157 \beta _1 C_F}{108\beta
   _0^2} +\frac{\beta
   _1^2}{6 \beta _0^3}-\frac{17 C_F}{27} \right)
        \right.
    \nonumber\\
   &&
   \left.     
   +
   \frac{n_f}{\beta _0}\left\{
  \left(\frac{26 \zeta (3)}{9}+\frac{4315}{648}\right) C_A C_F+\left(\frac{67}{27}-2 \zeta
   (3)\right) C_A^2+\left(\frac{11803}{2916}-\frac{8 \zeta (3)}{9}\right) C_F^2\right\}
      \right.
    \nonumber\\
   &&
   \left. 
   +\frac{61 C_A C_F}{108}-\frac{C_F^2}{27}\right]
   \frac{\left\langle N(p) \right| m \bar{\psi }\psi   \left| N(p) \right\rangle}{2M^2}
   \Biggr)-\frac{1}{4}A_q^{\rm NNLO}(\mu)
\ ,
\label{asy}
\end{eqnarray}
in the MS-like schemes,
with $\mu_0$ being a certain input scale.
Here, the first and the second lines show the LO terms that
are composed of the leading contributions from the terms proportinal to $x_q(\alpha_s)\frac{2g}{\beta (g)}$ in (\ref{qangan})
and of the terms of $A_q^{\rm LO}(\mu)$ given by (\ref{alo}); similarly, 
the third and the fourth lines 
show the NLO terms with $A_q^{\rm NLO}(\mu)$ given by
(\ref{anlo}), 
and the fifth and the following lines show the NNLO terms with $A_q^{\rm NNLO}(\mu)$ given by
(\ref{annlo}).
This is our main result that extends the two-loop 
calculation of \cite{Hatta:2018sqd} for constraining the twist-four gravitational form factor $\bar{C}_{q,g}$ into the  next higher-order accuracy.
As emphasized in \cite{Hatta:2018sqd}, the terms arising in the RHS of the first line are independent of the scale $\mu$
and represent the asymptotic value of $\bar{C}_q(\mu)=-\bar{C}_g(\mu)$
as $\mu \rightarrow \infty$; in the chiral limit, in particular, they are completely determined by the values of  $N_c$ and $n_f$,
as $\frac{1}{4}$ times the sum of  
\begin{equation}
-\frac{n_f}{4C_F+n_f}\ , \;\;\;\;\;\;\;\;\;\;   \mbox{and} \;\;\;\;\;\;\;\;\;\;\;\;\;\;\;   - \frac{2n_f}{3\beta_0}\ ,
\end{equation}
which come from the first term of (\ref{alo}) and the second term of (\ref{qanqanqan}), respectively,
and gives the values
\begin{equation}
-\frac{9}{25}\ \left(=-0.36\right)\ , \;\;\;\;\;\;\;\;\;\;   \mbox{and} \;\;\;\;\;\;\;\;\;\;\;\;\;\;\;    -\frac{2}{9}\ \left(=-0.22\ldots\right)\ ,
\end{equation}
for $N_c=3$, $n_f=3$ (compare with the first term of (\ref{ccc}) bellow).

As seen in (\ref{qangan}) with (\ref{zfzfzfh}), the $n$-loop terms (i.e., the order $\alpha_s^n$ terms) 
arising in $x_q(\alpha_s)$
contribute to $\bar{C}_{q,g}$ of (\ref{asy})
at order $\alpha_s^{n-1}$ and higher; this fact indicates that the naive counting in $\alpha_s$ does not work 
when deriving $\bar{C}_{q,g}$, as first pointed out at the two-loop level in \cite{Hatta:2018sqd}. 
Our three-loop result (\ref{asy}) shows that the corresponding $n$-loop level  
approximation for $\bar{C}_{q,g}$, retaining up to the order $\alpha_s^{n-1}$ terms, matches with 
the $\alpha_s$ counting in 
the $n$-loop level
results (\ref{aall})-(\ref{annlo}) for $A_{q,g}(\mu)$, corresponding to the N$^{n-1}$LO solution
of the RG equation, (\ref{one}), (\ref{one2}).
Therefore, (\ref{asy}) represents $\bar{C}_{q,g}$ that is organized according to the RG-improved perturbation theory
and is exact up to the corrections of N$^3$LO and higher.
Indeed, a key relation~(\ref{mass}) used to obtain (\ref{qangan}), (\ref{asy}) may be regarded as a consequence of solving the corresponding RG equations, because the RG-invariance relation, $\frac{d}{d\mu} T^\lambda_\lambda=0$, for the total trace anomaly (\ref{111}) is obeyed by (\ref{mass}) and yields the equation for the $\mu$ dependence of the operator $F^2$,
which is identical to the RG equation resulting from (\ref{o4ren0}), as demonstrated in \cite{Tanaka:2018nae}; 
the corresponding solution (\ref{massmass}), and thus the terms of (\ref{asy}) derived by its use, should obey the counting in $\alpha_s$ according to the RG-improved perturbation theory, as in (\ref{aall}).
In this context, it is also worth mentioning that the formula
(\ref{asy}) satisfies the RG equation, which is obtained as the matrix element 
of the 
three-loop evolution equation for the twist-four operator, 
\begin{eqnarray}
\frac{\partial}{\partial \ln \mu}
 \left(g\bar\psi F^{\lambda\nu}\gamma_\nu \psi \right)
& =&   
 \frac{\alpha_s}{4\pi}\left( 
 \left(-\frac{16 C_F}{3}-\frac{4 n_f}{3}\right) g\bar\psi F^{\lambda\nu}\gamma_\nu \psi +\frac{4 
   C_F}{3}\partial^\lambda\left( m \bar\psi \psi \right)
\right)
\nonumber\\&&
+\left(\frac{\alpha_s}{4\pi}\right)^2\left[
   \left(\frac{11 C_A}{18}+\frac{4 C_F}{9}\right) n_f \partial^\lambda F^2 
   \right.
   \nonumber\\
   &&
   +\left(
   \left(\frac{20 C_F}{9}-\frac{70 C_A}{27}\right)n_f-\frac{752 C_A C_F}{27}+\frac{224
   C_F^2}{27}\right)g\bar\psi F^{\lambda\nu}\gamma_\nu \psi  
  \nonumber\\
  &&\left.
    + \left(\frac{122 C_A C_F}{27}-\frac{136 C_F n_f}{27}-\frac{8
   C_F^2}{27}\right)\partial^\lambda\left( m \bar\psi \psi \right)
\right]
\nonumber
\\
&&
\!\!\!\!\!\!\!\!\!\!\!\!\!\!\!\!\!\!\!\!\!\!\!\!\!\!\!\!\!\!\!\!\!\!\!\!\!\!\!\!\!\!\!\!\!\!\!\!
+\left(\frac{\alpha_s}{4\pi}\right)^3\left[
\partial^\lambda F^2  \left(n_f^2 \left(-\frac{56 C_A}{81}-\frac{19 C_F}{27}\right)
+n_f
   \left(\frac{433 C_A C_F}{108}+\frac{1235 C_A^2}{324}+\frac{14
   C_F^2}{27}\right)\right)
   \right. \nonumber
\end{eqnarray}    
\begin{eqnarray}         
   &&\left.
   + g\bar\psi F^{\lambda\nu}\gamma_\nu \psi  \left(n_f \left(\left(16 \zeta
   (3)+\frac{322}{9}\right) C_A C_F+\left(48 \zeta (3)-\frac{3589}{81}\right)
   C_A^2
   \right.\right.\right. \nonumber
\\ 
   &&\left.\left.\left.
   +\left(\frac{9878}{243}-64 \zeta (3)\right) C_F^2\right)+n_f^2
   \left(\frac{2116 C_A}{243}+\frac{1358 C_F}{243}\right)+\left(128 \zeta
   (3)+\frac{17056}{243}\right) C_A C_F^2
   \right.\right. \nonumber
\\ 
   &&\left.\left.
   -\frac{16}{243} (648 \zeta (3)+2615)
   C_A^2 C_F-\frac{32}{243} (648 \zeta (3)-35) C_F^3\right)
   \right. \nonumber
\\ 
   &&\left.
   +\partial^\lambda\left( m \bar\psi \psi \right) \left(n_f
   \left(\left(\frac{64 \zeta (3)}{3}-\frac{8305}{243}\right) C_F^2-\frac{2}{81}
   (864 \zeta (3)+1079) C_A C_F\right)
   \right.\right. \nonumber\\ &&\left.\left.
   +\left(-32 \zeta
   (3)-\frac{1144}{243}\right) C_A C_F^2+\left(\frac{32 \zeta
   (3)}{3}+\frac{6611}{243}\right) C_A^2 C_F
    \right.\right. 
    \nonumber
\\ 
    &&\left.\left.
   -\frac{76}{81} C_F
   n_f^2+\frac{8}{243} (648 \zeta (3)-125) C_F^3\right)
\right]
\ .
\label{ee3}
\end{eqnarray}
Note that one has to take the off-forward matrix element of this evolution
equation to separate the overall factor $\Delta^\mu = p'^\mu -p^\mu$ as in (\ref{mat1}) when deriving the corresponding RG equation.
This evolution equation (\ref{ee3}) at the three-loop accuracy for the quark-gluon three-body operator of 
twist-four was derived in \cite{Tanaka:2018nae}.

\section{Calculating the nucleon's $\bar{C}_{q,g}$ at NNLO}
\label{secnucleon}

Our three-loop formula~(\ref{asy}) 
allows us to calculate $\bar{C}_{q,g}(\mu)$ as a function of the renormalization scale $\mu$
in the $\overline{\rm MS}$ scheme.
We calculate the values for $\bar{C}_{q,g}(\mu)$ at the NNLO accuracy for a proton,
using the coefficients of the beta-function (\ref{betaexact})
to three loops,
\begin{eqnarray}
\beta_0&=&\frac{11}{3}C_A-\frac{2n_f}{3}\ , 
\label{beta0}
\\
\beta_1&=&\frac{34}{3}C_A^2-2C_Fn_f -\frac{10}{3}C_An_f\ ,
\label{beta1}
\\
\beta_2&=&\frac{2857 C_A^3}{54}
-\frac{1}{2}
   n_f \left(\frac{1415 C_A^2}{27}+\frac{205 C_A C_F}{9}-2
   C_F^2\right)
+\frac{1}{4} n_f^2 \left(\frac{158 C_A}{27}+\frac{44 C_F}{9}\right)\ ,
\label{beta2}
\end{eqnarray}
and the corresponding three-loop running coupling constant, 
which is obtained by solving the RG equation of (\ref{betaexact}),
\begin{equation}
\frac{d\ln \alpha_s}{d\ln \mu^2}=\frac{\beta (g)}{2\sqrt{\pi \alpha_s}}=-\beta_0 \frac{\alpha_s}{4\pi}-\beta_1 \left(\frac{\alpha_s}{4\pi}\right)^2
-\beta_2 \left(\frac{\alpha_s}{4\pi}\right)^3\ ,
\label{run}
\end{equation}
as
\begin{eqnarray}
\ln \frac{\mu^2}{\Lambda_{\rm QCD}^2}=
\frac{4
   \pi }{\beta _0 \alpha _s(\mu )}+\frac{\beta _1 }{\beta _0^2}\ln \left(\frac{\beta _0 \alpha _s(\mu
   )}{4 \pi }\right)
+
\frac{\left(\beta _0 \beta _2-\beta _1^2\right) \alpha _s(\mu )}{4 \pi  \beta _0^3}
\ .
\label{lambdaQCD}
\end{eqnarray}
Here, the constant of integration is represented by the QCD scale parameter $\Lambda_{\rm QCD}$ according to the definition in \cite{Chetyrkin:2000yt,Collins:2011zzd}; although (\ref{lambdaQCD})
may be further solved for $\alpha _s(\mu )$ iteratively, leading to
\begin{equation}
\frac{\alpha _s(\mu )}{4\pi}=\frac{1}{\beta _0 L}
-\frac{\beta _1 \ln L}{\beta _0^3 L^2}
+\frac{1}{L^3}\left[\frac{\beta _2}{\beta _0^4}+\frac{\beta _1^2 \left(\ln ^2L-\ln L-1\right)}{\beta
   _0^5}\right]
\ ,
\label{running}
 \end{equation}
with $L\equiv \ln \left(\mu^2/\Lambda_{\rm QCD}^2\right)$, we shall use
the exact numerical solution of  (\ref{lambdaQCD}) 
as the value of $\alpha _s(\mu )$ in our calculations.

We evaluate (\ref{asy}) with (\ref{anlo}) and (\ref{annlo}), assuming a fixed number $n_f=3$.
Substituting $N_c=3$ and $n_f=3$ into (\ref{asy}), we obtain
\begin{eqnarray}
{\left. {\bar C_q(\mu )} \right|_{{n_f} = 3}} &&= -0.145556+0.305556 \frac{\left\langle N(p) \right| m \bar{\psi }\psi   \left| N(p) \right\rangle}{2M^2}+
 \left(0.09\, -0.25 A_q\left(\mu
   _0\right)\right)\left(\frac{\alpha _s\left(\mu \right)}{\alpha _s\left(\mu _0\right)}\right)^\frac{50}{81}
\nonumber\\
&& 
+\alpha _s(\mu ) \Biggl[0.00553609+0.0803962 \frac{\left\langle N(p) \right| m \bar{\psi }\psi   \left| N(p) \right\rangle}{2M^2}
\nonumber\\
&& \;\;\;\;
+
   \left(0.0127684\, -0.0354678 A_q\left(\mu _0\right)\right)\left(\frac{\alpha _s\left(\mu \right)}{\alpha _s\left(\mu _0\right)}\right)^\frac{50}{81}
\nonumber\\
&& \;\;\;\;
-\left(0.0279651-0.0354678 A_q\left(\mu
   _0\right)\right)\left(\frac{\alpha _s\left(\mu \right)}{\alpha _s\left(\mu
   _0\right)}\right)^{-\frac{31}{81}}
\Biggr]
\nonumber
\end{eqnarray}
\begin{eqnarray}
&&
+
\bigl(\alpha _s(\mu )\bigr)^2 \Biggl[0.00174426+0.0312256 \frac{\left\langle N(p) \right| m \bar{\psi }\psi   \left| N(p) \right\rangle}{2M^2}
\nonumber\\
&&   \;\;\;\;
   -\left(0.0059729-0.0165914 A_q\left(\mu
   _0\right)\right) \left(\frac{\alpha _s\left(\mu \right)}{\alpha _s\left(\mu
   _0\right)}\right)^\frac{50}{81} 
      \nonumber\\
&&\;\;\;\;
   -\left(0.00396745-0.00503187 A_q\left(\mu _0\right)\right)\left(\frac{\alpha _s\left(\mu
  \right)}{\alpha _s\left(\mu _0\right)}\right)^{-\frac{31}{81}}
\nonumber\\
&&\;\;\;\;
+
\left(0.0237481\, -0.0216233 A_q\left(\mu _0\right)\right)\left(\frac{\alpha _s\left(\mu
   \right)}{\alpha _s\left(\mu _0\right)}\right)^{-\frac{112}{81}}
  \Biggr]
\ ,
\label{ccc}
\end{eqnarray}
up to the corrections of N$^3$LO and higher.
Here, for $\alpha _s(\mu )$, we use the value determined by (\ref{lambdaQCD}) with  $\Lambda_{\rm QCD}^{(3)}\simeq0.3359$~GeV,
so that
\begin{equation}
\alpha_s(\mu=1~\rm{GeV})\simeq 0.4736\ .\footnote{
This value would evolve into the conventional
$\alpha_s(M_Z)=0.1181$, if
the number of active flavors were determined automatically
such that the decoupling is 
performed at the pole mass of the respective heavy quark,
using the RunDec package~\cite{Herren:2017osy}.
In the present case with $n_f=3$ fixed, we have $\alpha_s(M_Z)\simeq 0.1059$.}
\label{alphas1gev}
\end{equation}
The corresponding NNLO coupling constant by (\ref{lambdaQCD})
is always used 
in the following numerical computations, independently of the order considered, as a way of isolating the effect
of the higher order contributions exhibited in the formula~(\ref{ccc}).
For the values of the input scale $\mu_0$ and the value for $A_q\left(\mu _0\right)$, we use
\begin{equation}
\mu_0=1.3~{\rm GeV}\ ,\;\;\;\;\;\;\;\;
A_q\left(\mu _0=1.3~{\rm GeV}\right)=0.613\ ,
\label{inputs}
\end{equation}
These values correspond to the starting scale and the total momentum fraction shared by the three quark flavors, $u$, $d$ and $s$
in the CT18 parton distribution functions of the nucleon~\cite{Hou:2019efy},
which are determined by the global QCD analysis at NNLO; at the starting scale $\mu_0=1.3~{\rm GeV}$ of CT18,
the active quark flavors arising in (\ref{aqmu}) are $u, d$ and $s$.
We note that, for  (\ref{inputs}), the uncertainty $\lesssim$ a few percent~\cite{Hou:2019efy}, and 
it is consistent with the results of the
other collaborations of the 
global QCD analysis like \cite{Harland-Lang:2014zoa,NNPDF:2017mvq} within such small uncertainties. 

\begin{figure}[hbtp]
\begin{center}
\includegraphics[width=0.47\textwidth]{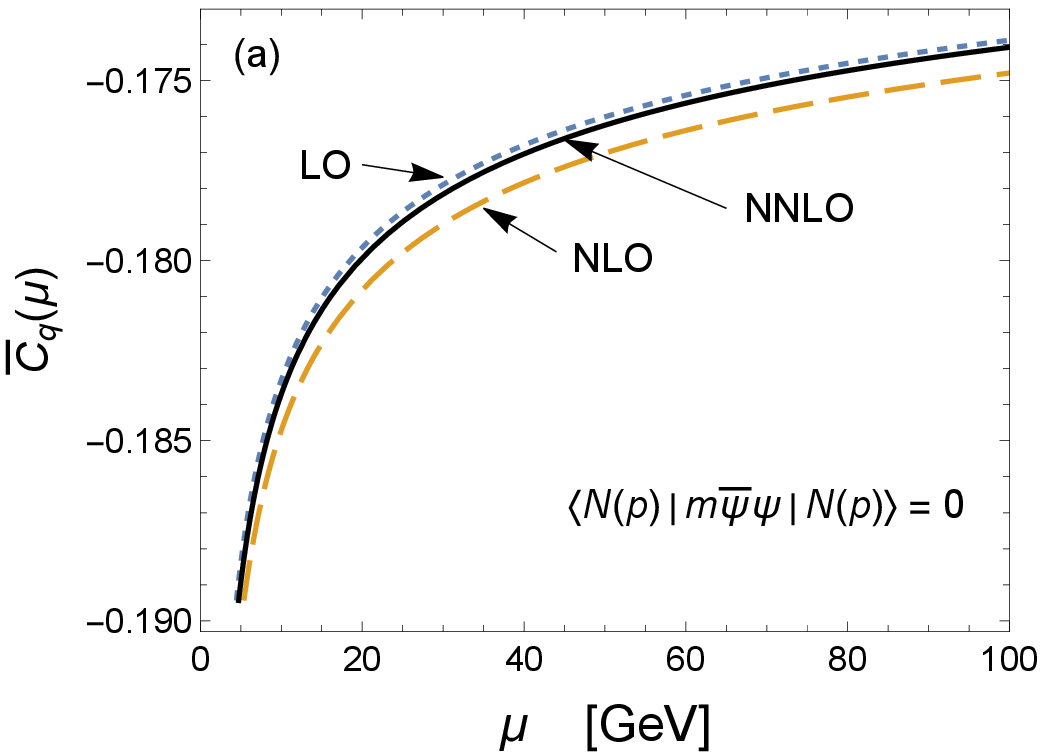}
\hspace{0.5cm}
\includegraphics[width=0.47\textwidth]{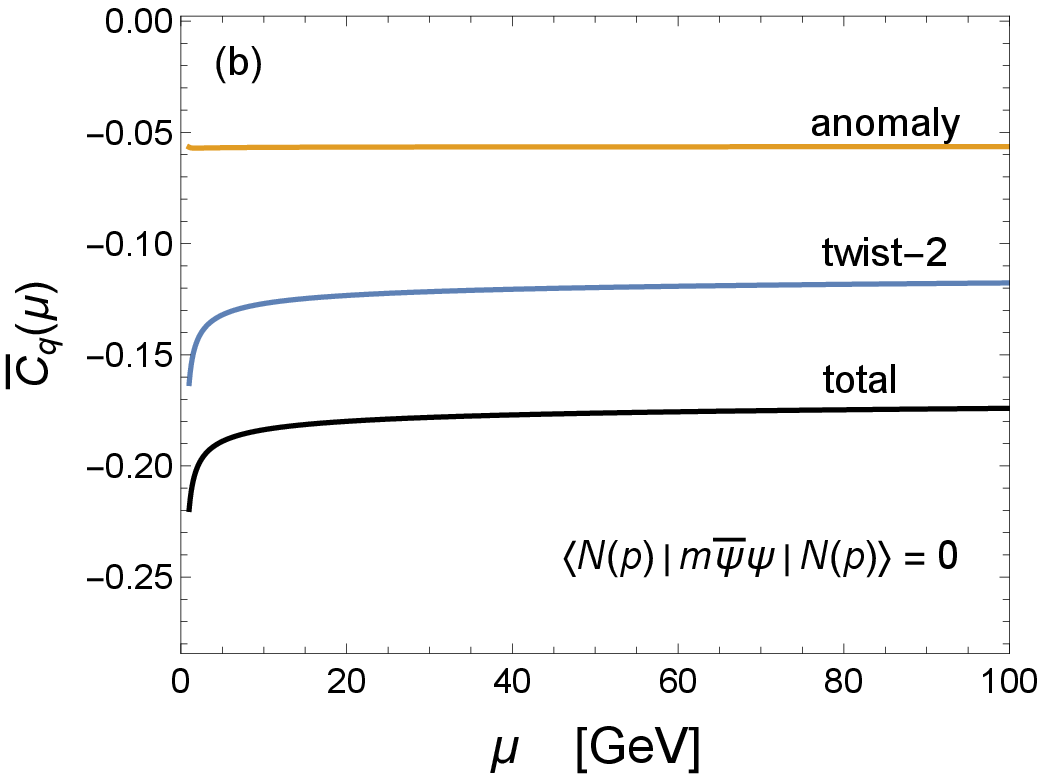}
\caption{The 
nucleon's gravitational form factor $\left. {\bar C_q(\mu )} \right|_{{n_f} = 3}$ of (\ref{ccc}) at the NNLO (3-loop) accuracy
in the chiral limit, $\langle N (p)|m \bar \psi \psi| N (p)\rangle=0$:
(a) the results up to the LO, NLO, and NNLO contributions; (b) the total (NNLO) result, and the separate contributions from the first (twist-2 effect) and second (anomaly effect) terms of (\ref{trace}).
}
\label{fig1}
\end{center}
\end{figure}

Firstly, (\ref{ccc}) in the chiral limit is plotted as a function of $\mu$ in Fig.~\ref{fig1}: Fig.~\ref{fig1}(a) shows the results up to the LO, NLO, and NNLO contributions;
the NLO as well as NNLO corrections give a
few percent level effects, reflecting the small numerical coefficients 
for those correction terms arising in (\ref{ccc}),
and, furthermore,
the NLO and NNLO corrections tend to cancel.
Thus, the important correction comes from the LO-level evolution of the twist-two form factor $A_q$,
so that the approach to the asymptotic value ($\simeq -0.146$) is quite slow,
while the other corrections play a minor ($\lesssim$ a few percent) role.   
In Fig.~\ref{fig1}(b), the NNLO result is separated into the individual contributions from each term in (\ref{trace}), the first (twist-2 effect) term and the second (anomaly effect) term; both twist-2 and anomaly effects produce the important contributions.

When taking into account the quark-mass effects in evaluating (\ref{ccc}),
we need the matrix element of the quark scalar operator,  $\langle N (p)|m \bar \psi \psi| N (p)\rangle$, 
which is related to 
the sigma terms (see, e.g., \cite{Gasser:1982ap,Hatsuda:1994pi,Ji:1994av,Meissner:2022odx}).
Assuming $n_f=3$, the corresponding relation reads 
\begin{equation}
\langle N (p)|m \bar \psi \psi| N (p)\rangle=\langle N (p)|m_u \bar u u+m_d \bar d d+m_s \bar s s | N (p)\rangle
=2M\left(\sigma_{\pi N}+ \sigma_{s}\right)\ ,
\label{sigmaterms}
\end{equation}
up to small isospin-violating corrections of ${\cal O}\left(m_d-m_u \right)$,
where
\[ \sigma_{\pi N}=\frac{1}{2M}\langle N (p)|\frac{m_u+m_d}{2}\left( \bar u u+ \bar d d\right) | N (p)\rangle
\]
is the pion-nucleon sigma-term, and 
\[ \sigma_{s}=\frac{1}{2M}\langle N (p)|m_s \bar s s 
 | N (p)\rangle
\] 
is the strangeness content of the nucleon.
Here, for the former, we use the value due to a recent phenomenological analysis~\cite{Hoferichter:2015dsa}, 
\begin{equation}
\sigma_{\pi N}=59.1\pm 3.5~{\rm MeV}\ ,
\label{emp1}
\end{equation}
and, for the latter, we use the value,
\begin{equation}
\sigma_{s}=45.6\pm 6.2~{\rm MeV}\ ,
\label{emp2}
\end{equation}
which is given by a recent lattice QCD determination~\cite{Alexandrou:2019brg}; see also \cite{Alarcon:2011zs,RuizdeElvira:2017stg,Hoferichter:2015hva,Alarcon:2012nr,Junnarkar:2013ac,Ren:2012aj,Ren:2014vea,Ren:2017fbv,Gupta:2021ahb,Yang:2015uis,Yamanaka:2018uud,XQCD:2013odc,FlavourLatticeAveragingGroupFLAG:2021npn}.

\begin{figure}[hbtp]
\begin{center}
\includegraphics[width=0.47\textwidth]{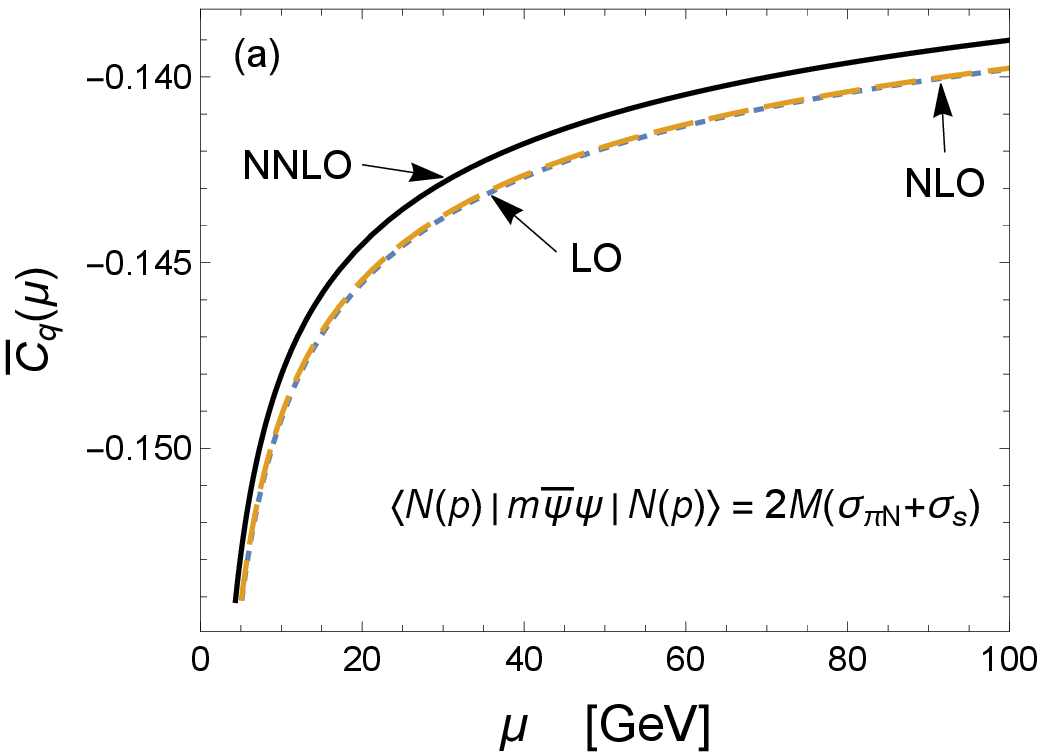}
\hspace{0.5cm}
\includegraphics[width=0.47\textwidth]{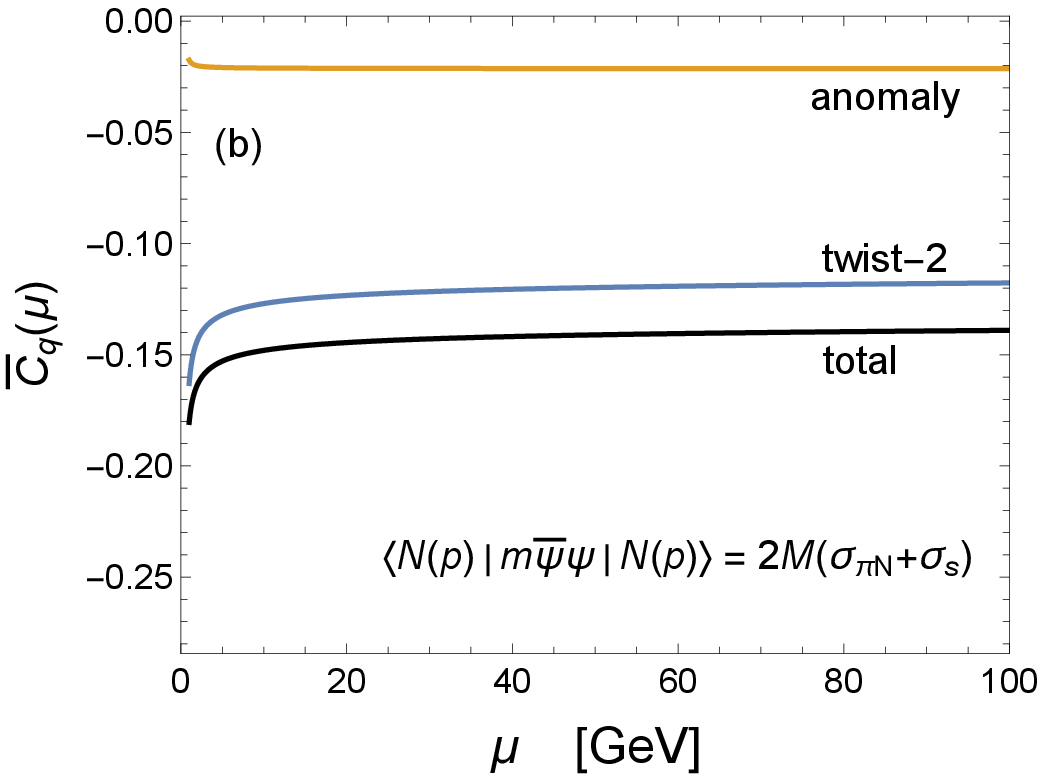}
\caption{The 
nucleon's gravitational form factor $\left. {\bar C_q(\mu )} \right|_{{n_f} = 3}$ of (\ref{ccc}) at the NNLO (3-loop) accuracy
with $\langle N (p)|m \bar \psi \psi| N (p)\rangle
=2M\left(\sigma_{\pi N}+ \sigma_{s}\right)$, using (\ref{emp1}), (\ref{emp2}):
(a) the results up to the LO, NLO, and NNLO contributions; (b) the total (NNLO) result, and the separate contributions from the first (twist-2 effect) and second (anomaly effect) terms of (\ref{trace}).
}
\label{fig2}
\end{center}
\end{figure}

Fig.~\ref{fig2} shows (\ref{ccc}) with (\ref{sigmaterms})-(\ref{emp2}) as a function of $\mu$, similarly as Fig.~\ref{fig1}.
Compared to Fig.~\ref{fig1}(a) in the chiral limit, the sigma terms increase the LO value of $\left. {\bar C_q(\mu )} \right|_{{n_f} = 3}$ in Fig.~\ref{fig2}(a) by $\sim 20$~\% due to the contribution from the second term of (\ref{ccc}).
The sigma terms at the NLO level, due to the second term in the second line of  (\ref{ccc}), also give the positive
contribution and almost cancel the negative contribution due to the NLO evolution arising in Fig.~\ref{fig1}(a),
so that the LO and the NLO curves are indistinguishable in Fig.~\ref{fig2}(a).
Thus, the NLO as well as NNLO terms of (\ref{ccc}) give at most a
few percent level effects, again,
reflecting the small numerical coefficients 
of the corresponding terms in (\ref{ccc}).
Fig.~\ref{fig2}(b) also demonstrates that the sigma terms give positive effects to the anomaly contribution,
compared to the results in Fig.~\ref{fig1}(b).

\begin{figure}[hbtp]
\begin{center}
\includegraphics[width=0.5\textwidth]{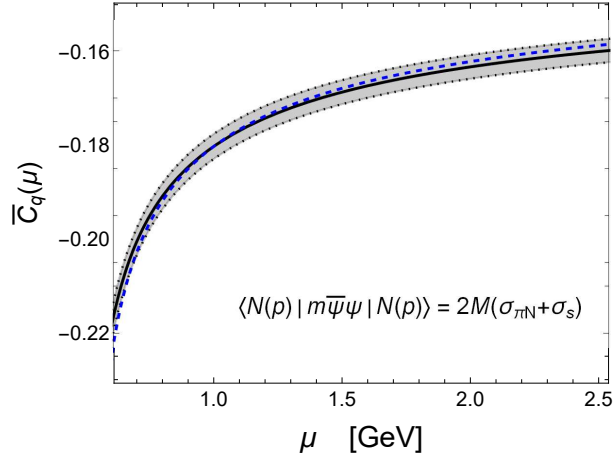}
\caption{The 
nucleon's gravitational form factor $\left. {\bar C_q(\mu )} \right|_{{n_f} = 3}$ of (\ref{ccc}) at the NNLO (3-loop) accuracy
with $\langle N (p)|m \bar \psi \psi| N (p)\rangle
=2M\left(\sigma_{\pi N}+ \sigma_{s}\right)$: the solid line shows the full (NNLO) result; the shaded areas indicate the uncertainties estimated by varying the sigma terms within
the uncertainties of  (\ref{emp1}), (\ref{emp2}).
The blue dashed line shows the approximate formula (\ref{approxf}).
}
\label{fig3}
\end{center}
\end{figure}

The sigma terms (\ref{sigmaterms})-(\ref{emp2}) modify the asymptotic value due to the first two terms of (\ref{ccc}) 
into the value as
\begin{equation}
-0.145556+0.305556 \frac{\left\langle N(p) \right| m \bar{\psi }\psi   \left| N(p) \right\rangle}{2M^2}\simeq -0.111\ .
\label{asymptoticv}
\end{equation}
Although the approach to this asymptotic value is 
quite slow in Fig.~\ref{fig2}(a), similarly as in Fig.~\ref{fig1}(a), this value dominantly determines the size of 
$\left. {\bar C_q(\mu )} \right|_{{n_f} = 3}$
and the sigma terms contribute to (\ref{asymptoticv}) by only $\sim 20$~\%. As a result, rather large uncertainties of 
the sigma terms (\ref{sigmaterms})-(\ref{emp2}) do not cause large errors in $\left. {\bar C_q(\mu )} \right|_{{n_f} = 3}$:
the solid curve in Fig.~\ref{fig3} shows our NNLO formula~(\ref{ccc}) with (\ref{sigmaterms})-(\ref{emp2}) as a function of $\mu$ for the
region~\footnote{We note that the corresponding evolution is performed 
not
only to the scales higher than the input scale $\mu_0=1.3$~GeV, but also 
to the scales lower than $\mu_0=1.3$~GeV; the inclusion of the high orders of perturbation theory
for our evolution equations allows a reliable evaluation even for the “backward” evolution towards the low scales 
(see e.g., \cite{Altenbuchinger:2010sz,deFlorian:2019egz} for a backwards evolution in a different context).} suitable for the present evaluation assuming a fixed number $n_f=3$;
here, 
the shaded areas display the uncertainties estimated by varying the sigma terms within
the uncertainties of (\ref{emp1}), (\ref{emp2}).
Some of the corresponding explicit values read
\begin{eqnarray}
\left. {\bar C_q(\mu=0.7~{\rm GeV} )} \right|_{{n_f} = 3}&=&-0.201\pm 0.003\ ,
\nonumber\\
\left. {\bar C_q(\mu=1~{\rm GeV} )} \right|_{{n_f} = 3}&=&-0.180\pm 0.003\ ,
\nonumber\\
\left. {\bar C_q(\mu=2~{\rm GeV} )} \right|_{{n_f} = 3}&=&-0.163\pm 0.003\ .
\label{explicitresults}
\end{eqnarray}
These are the values in the $\overline{\rm MS}$ scheme.
As discussed in Figs.~\ref{fig1}(a) and \ref{fig2}(a) above, the NNLO terms in (\ref{ccc}) produce a few \% level
effects. Thus, we believe that the uncertainties due to the omission of the terms of N$^3$LO and higher 
should be much smaller than the uncertanties presented in Fig.~\ref{fig3} and (\ref{explicitresults}).\footnote{This fact
may be explicitly checked using a recent extension of the quark/gluon decomposition of the trace 
anomaly~(\ref{tgrenaq}), (\ref{tgrenag}) to the four-loop order~\cite{Ahmed:2022adh}, but we do not go into the detail here.} 
It is remarkable that, although the uncertainties in the inputs from the sigma terms (\ref{emp1}), (\ref{emp2})
are rather large and determine the uncertainties in the final results, 
the small numerical coefficients associated with the sigma terms 
in (\ref{ccc}) lead to the resultant uncertainties at a few \% level, allowing us 
to obtain the accurate predictions as in Fig.~\ref{fig3} and (\ref{explicitresults})
without spoiling the accuracy of the perturbative calculations at the NNLO.
We note that an estimate in \cite{Hatta:2018sqd} using the asymptotic value as $\left. \bar{C}_q\right|_{{n_f} = 3} \approx -0.15$
is confirmed and improved by the present results~(\ref{explicitresults}),
but the values of (\ref{explicitresults}) are rather different from those of other estimates~\cite{Ji:1997gm,Polyakov:2018exb,Lorce:2018egm,Azizi:2019ytx}.

The approach of the curves to the asymptotic limit (\ref{asymptoticv}) in Fig.~\ref{fig2}(a) as well as in Fig.~\ref{fig1}(a) 
is slow as a function of $\mu$, because it is controlled by the logarithm of $\mu$ as in (\ref{running}).
We find in the calculations of those figures that the values different from the asymptotic limit by $10$~\% or less are obtained 
for the huge $\mu$ like $\mu \gtrsim 10^{12}$~{\rm GeV},
which corresponds to
\beq
 \ln \left(\mu/\Lambda_{\rm QCD}\right)
\gtrsim  \frac{2\pi}{\beta_0}10^{\frac{3 \beta_0}{ 8 C_F+2n_f}}\ ,
\eeq
so that
$\left[\alpha _s\left(\mu
   \right)\right]^{(8 C_F+2n_f)/3 \beta_0}
\lesssim 1/10$, see (\ref{asy}).
Therefore, the asymptotic value (\ref{asymptoticv}) should be regarded as the formal $\mu \to \infty$ limit 
of our formulas (\ref{asy}), (\ref{ccc}), and does not represent the leading contribution in any sense in quantitative evaluations.
Indeed,  according to the order counting explained below (\ref{asy}), the leading order contribution is composed of the asymptotic value (\ref{asymptoticv})
and the LO evolution contribution (i.e., the second line of  (\ref{asy})),
such that the latter is as important as the former.

The behaviors of the NLO and NNLO contributions observed in Figs.~\ref{fig1}, \ref{fig2}
suggest that the LO terms in the first line of (\ref{ccc}) with the asymptotic value (\ref{asymptoticv}) modified into the 
corresponding NNLO value $\simeq -0.108$ could provide a good approximation of the full NNLO result.
The correspnding approximation reads
\begin{equation}
\left. {\bar C_q(\mu )} \right|_{{n_f} = 3} \simeq -0.108
-0.114\left[\alpha _s\left(\mu \right)\right]^\frac{50}{81}\ ,
\label{approxf}
\end{equation}
where the second term coincides with the LO evolution term (the third term  of (\ref{ccc})) with (\ref{inputs}) substituted,
and this approximate formula is plotted by the blue dashed curve in Fig.~\ref{fig3}. 

The gravitational form factors of (\ref{para}) are studied in lattice QCD calculations~\cite{Shanahan:2018nnv,Shanahan:2018pib},
but the corresponding calculation of the twist-four gravitational factor $\bar{C}_{q,g}(t, \mu)$ seems to be still missing.
Recently the behaviors of $\bar{C}_{q,g}(t, \mu)$ are studied with perturbative QCD factorization~\cite{Tong:2021ctu,Tong:2022zax}, but this framework is 
applicable to the cases with large momentum transfer $t$.
We emphasize that the NNLO QCD prediction of the forward value $\bar{C}_{q,g}(0, \mu)$ is now available.
As presented above, the quark/gluon decomposition of the QCD trace anomaly (\ref{tgrenaq}), (\ref{tgrenag}) provide
sufficient constraints to allow us to obtain a model-independent determination of the forward value as (\ref{explicitresults}), (\ref{approxf}), up to a few \% uncertainties.

\section{Spin-0 hadron case}
\label{pionsec}

The matrix element of the quark part of the energy-momentum tensor of (\ref{tg}) in terms of  a spin-0 hadron state, 
$|h(p)\rangle$, like a pion state, is parameterized as (see e.g. \cite{Tanaka:2018wea,Tong:2021ctu})
\begin{eqnarray}
\!\!\!
\!\!\!
\langle h( p')|T_{q}^{\mu \nu}|h(p)\rangle 
=\frac{1}{2}{\Theta}_{2q}(t) \bar{P}^\mu\bar{P}^\nu+
\frac{1}{2}{\Theta}_{1q}(t)\left(tg^{\mu\nu}- \Delta^\mu \Delta^\nu\right)+  2m_h^2\bar{C}^h_q(t)\eta^{\mu \nu}\ ,
\label{tmunumatred}
\end{eqnarray}
where
$m_h$ denotes the mass of the hadron $h$,
and the matrix element of the gluon part of (\ref{tg}) is given by the similar parameterization with $q\to g$. 
The dimensionless Lorentz-invariant coefficients, ${\Theta}_{1q}(t), {\Theta}_{2q}(t), \bar{C}^h_q(t)$, ${\Theta}_{1g}(t), {\Theta}_{2g}(t), \bar{C}^h_g(t)$, are the gravitational form factors for a spin-0 hadron $h$.
Similar to (\ref{camu}), we treat the form factors relevant to the forward limit, as 
\begin{eqnarray}
&&\bar{C}^h_{q,g}(\mu) \equiv \bar{C}^h_{q,g}(t=0, \mu)\ , \nonumber\\
&&{\Theta}_{2q}(\mu)\equiv {\Theta}_{2q}\left(t=0, \mu\right)\ ,\;\;\;\;\;\;\;{\Theta}_{2g}(\mu)\equiv {\Theta}_{2g}\left(t=0, \mu\right)\ ,
\label{camuh}
\end{eqnarray}
denoting the renormalization scale $\mu$ dependence explicitly in the following.

It is straightforward to see that the manipulations with (\ref{tmunumatred}), (\ref{camuh}), similar as in Sec.~\ref{sec4},
lead to the formulas (\ref{aqag1})-(\ref{asy}) with the substitutions, 
\begin{eqnarray}
&&M \to m_h\ , \;\;\;\;\; \;\;\;\;\; \;\;\;\;\; \;\;\;\;\; \;
\left| N(p) \right\rangle \to \left| h(p) \right\rangle\ ,
\nonumber\\
&&\bar{C}_j(\mu)\to 
\bar{C}^h_j(\mu)\ ,\;\;\;\;\; \;\;\;\;\;
A_j(\mu) \to \frac{1}{4}{\Theta}_{2j}(\mu)\ ,
\label{substitution}
\end{eqnarray}
with $j=q, g$; e.g., (\ref{qangan}) with these substitutions
read
\begin{eqnarray}
\bar{C}^h_q(\mu)
&&= \frac{1}{4}\left( - \frac{1}{4}{\Theta}_{2q}(\mu)+   x_q(\alpha_s)\frac{2g}{\beta (g)}
\right)
\nonumber\\
&&
+ \left\{1+y_q(\alpha_s)-x_q(\alpha_s)\frac{2g}{\beta (g)} \left(1+\gamma_m(g)\right)\right\}\frac{\langle h(p)|m\bar{\psi}\psi|h(p)\rangle}{8m_h^2}
\ ,
\label{qanganpi}
\end{eqnarray}
and,
combining this with the evolution (\ref{oneintegratea}) with the substitutions $A_j \to {\Theta}_{2j}$,
we can determine the value of $\bar{C}^h_{q,g}(\mu)$ of a spin-0 hadron for arbitrary $\mu$
to the desired accuracy.
As a result, $\bar{C}^h_{q,g}(\mu)$ at the NNLO accuracy in the MS-like schemes are expressed as (\ref{asy}) with the substitutions~(\ref{substitution}).

Among the spin-0 hadrons, the pion is of special interest\footnote{The behaviors of the gravitational form factors for the pion 
have been obtained~\cite{Kumano:2017lhr} 
through the determination of the generalized distribution amplitudes (GDAs)~\cite{Diehl:1998dk,Diehl:2000uv,Diehl:2003ny,Kawamura:2013wfa}
using the Belle data on $\gamma^* \gamma  \rightarrow \pi^0 \pi^0 $.};
here, we evaluate (\ref{qanganpi}) for the case with the pion, $h=\pi$,
taking into account nontrivial nature as a Nambu-Goldstone boson.
The PCAC relation ($f_\pi$ is the pion decay constant),
\begin{equation}
- \left(m_u + m_d\right)\langle 0|\bar{u}u +\bar d d  |0\rangle = 2f_\pi^2m_\pi^2 \ ,
\end{equation}
due to Gell-Mann, Oakes, and Renner~\cite{Gell-Mann:1968hlm},
indicates $m_\pi^2 \sim m$ as $m\to 0$; therefore, even in the chiral limit,
we cannot neglect 
the terms associated with $\bigl\langle h(p)\bigl| m\bar{\psi} \psi
 \bigr|h(p)\bigr\rangle/m_h^2$ in (\ref{qanganpi}) for $h=\pi$. 
By contrast to the nucleon case discussed in Sec.\ref{secnucleon}, however,
it is remarkable that the corresponding matrix element, $\bigl\langle \pi (p)\bigl| m\bar{\psi} \psi
 \bigr|\pi (p)\bigr\rangle$, can be determined reflecting the Nambu-Goldstone nature of the pion.
We note that the pion mass can be calculated as the mass shift from the chiral limit, due to the ordinary
first-order perturbation theory in the quark mass term in the QCD Hamiltonian,
as~\cite{Gasser:1980sb,Gasser:1982ap,Meissner:2022odx}
\begin{equation}
m_\pi^2=\ _{0}\bigl\langle \pi(p)\bigl| m\bar{\psi} \psi
 \bigr|\pi(p)\bigr\rangle_0\ ,
\label{chiralp}
\end{equation}
where $\bigr|\pi(p)\bigr\rangle_0\equiv  \left. \bigr|\pi(p)\bigr\rangle\right|_{m=0}$,
so that we obtain 
\begin{equation}
\frac{\langle \pi(p)|m\bar{\psi}\psi|\pi(p)\rangle}{m_\pi^2} =1\ ,
\label{chiralpp}
\end{equation}
up to the corrections of ${\cal O}(m)$. It is worth mentioning that the matrix element $\bigl\langle \pi(p)\bigl|  F^2\bigr|\pi(p)\bigr\rangle$ is
also expressed by the pion mass, as
\begin{eqnarray}
\frac{\bigl\langle \pi(p)\bigl|  F^2\bigr|\pi(p)\bigr\rangle}{m_\pi^2}
=
\frac{2g}{\beta (g)}\left(1-\gamma_m(g)\right)
\ ,
\label{masspi3}
\end{eqnarray}
up to the corrections of ${\cal O}(m)$, using (\ref{chiralpp}) in (\ref{massmass}) with the substitutions (\ref{substitution}),
and that the relations (\ref{chiralpp}) and (\ref{masspi3}) have been utilized to determine
the anomaly-induced mass structure of the pion in \cite{Tanaka:2018nae}.
Using (\ref{chiralpp}), the above result (\ref{qanganpi}) with $h=\pi$ reads
\begin{equation}
\bar{C}^\pi_q(\mu)
= \frac{1}{8}\left( - \frac{1}{2}{\Theta}_{2q}(\mu)+x_q(\alpha_s)\frac{2g}{\beta (g)} \left(1-\gamma_m(g)
\right)
+ 1+y_q(\alpha_s)\right)
\ ,
\label{qanganpipipi}
\end{equation}
up to the corrections of ${\cal O}(m)$.
This result leads to the explicit NNLO-level formula 
for the pion's twist-four gravitational form factor $\bar{C}^\pi_q(\mu)=-\bar{C}^\pi_g(\mu)$
in the MS-like schemes,
which is given by (\ref{asy})
with the substitutions (\ref{substitution}), and also with $h=\pi$ and (\ref{chiralpp}) substituted;
in the $\overline{\rm MS}$ scheme and 
with $N_c=3$ and a fixed number of quark flavors $n_f=3$,
$\left. \bar{C}^\pi_q(\mu)\right|_{{n_f} = 3}$ 
is given at the NNLO accuracy by the formula (\ref{ccc}) with the replacements,
\begin{eqnarray}
&&A_q(\mu_0) \to \frac{1}{4}{\Theta}_{2q}(\mu_0)\ ,
\nonumber\\
&&\frac{\left\langle N(p) \right| m \bar{\psi }\psi   \left| N(p) \right\rangle}{2M^2} \to \frac{1}{2}\ .
\label{replacenpi}
\end{eqnarray}
Similar as (\ref{aqmu}) and (\ref{xfmu}), we have the relation (see (\ref{substitution})),
\begin{equation}
\frac{1}{4}{\Theta}_{2q}(\mu)= \sum_f \int_0^1dx x \left(q_f^\pi(x, \mu)+q^\pi_{\bar{f}}(x, \mu)\right)\ ,
\label{thetaqmumoment}
\end{equation}
using the quark and antiquark distribution functions for a pion, $q_f^\pi(x, \mu)$ and $q^\pi_{\bar{f}}(x, \mu)$, 
and the recent NLO global QCD analyses for those distribution functions, 
such that the active quark flavors at the scale $\mu_0=1.3$~GeV being $u, d$ and $s$, 
give
\begin{equation}
\frac{1}{4}{\Theta}_{2q}\left(\mu _0\right)=
\begin{cases}
0.70\pm 0.02 &\left(\mbox{Ref.\cite{Barry:2018ort}}\right)\ ,\\
0.81\pm 0.16 &\left(\mbox{Ref.\cite{Novikov:2020snp}}\right)\ ,\\
 0.61\pm 0.08 &\left(\mbox{Ref.\cite{Barry:2021osv}}\right)\ ,
\end{cases}
\label{theta2qinput}
\end{equation}
where the last analysis of \cite{Barry:2021osv} takes into account also the next-to-leading logarithmic threshold resummation on 
the relevant Drell-Yan cross sections,
which tends to make the valence distribution considerably softer at high momentum fractions $x$~\cite{Aicher:2010cb}. 
See also \cite{Sutton:1991ay,Gluck:1991ey,Gluck:1999xe,Wijesooriya:2005ir} 
for earlier works of the NLO QCD analysis and 
\cite{ExtendedTwistedMass:2021rdx} for a recent lattice result.
\begin{figure}[hbtp]
\begin{center}
\includegraphics[width=0.47\textwidth]{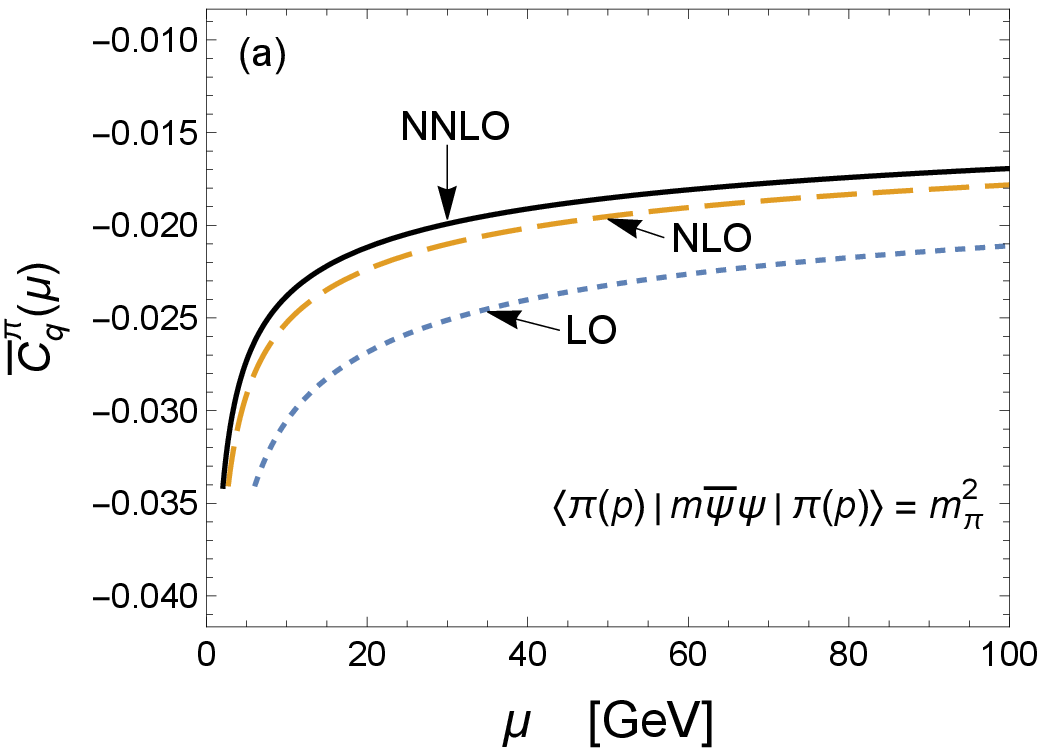}
\hspace{0.5cm}
\includegraphics[width=0.47\textwidth]{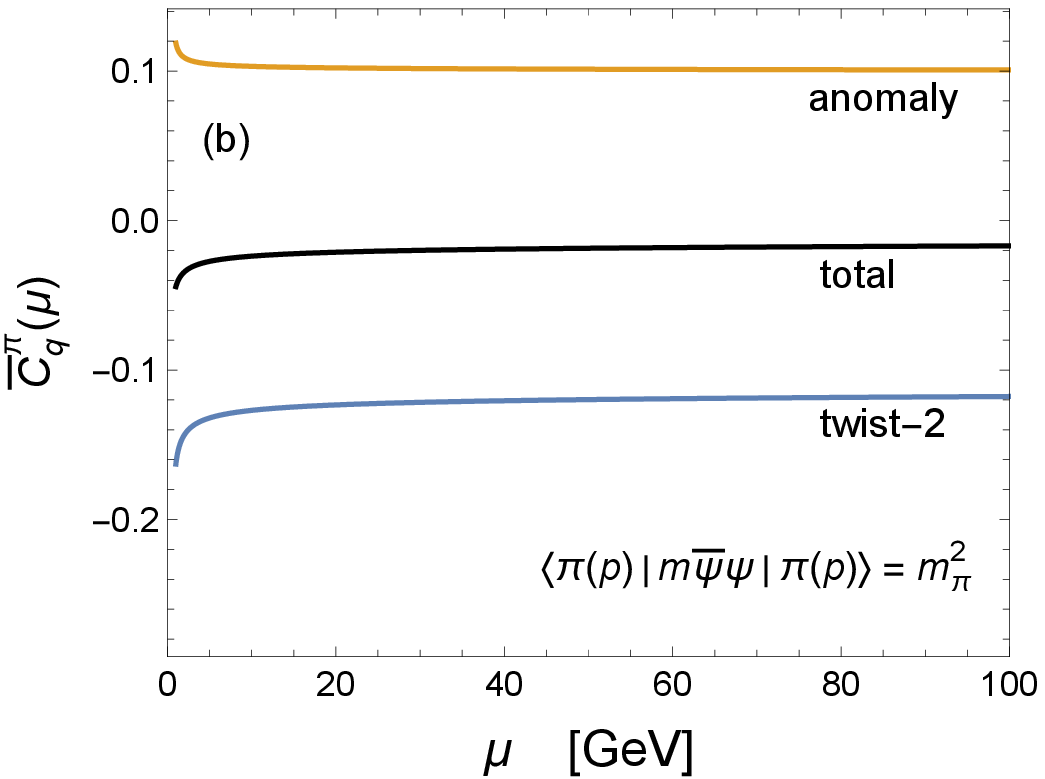}
\caption{The 
pion's gravitational form factor $\left. {\bar C}^\pi_q(\mu )\right|_{n_f=3}$ at the NNLO (3-loop) accuracy,
given by (\ref{ccc}) with (\ref{replacenpi}) and $\frac{1}{4}{\Theta}_{2q}\left(\mu _0=1.3~{\rm GeV}\right)=0.613$:
(a) the results up to the LO, NLO, and NNLO contributions; (b) the total (NNLO) result, and the separate contributions from the first (twist-2 effect) and the remaining (anomaly effect) terms of (\ref{qanganpipipi}).
}
\label{figpi1}
\end{center}
\end{figure}
First of all, we plot the above result of $\left. \bar{C}^\pi_q(\mu)\right|_{{n_f} = 3}$ 
in Fig.~\ref{figpi1} using  
$\frac{1}{4}{\Theta}_{2q}\left(\mu _0=1.3~{\rm GeV}\right)=0.613$, which produces the same contributions through
evolutions as those in Figs.\ref{fig1}, \ref{fig2} with the input (\ref{inputs}): Fig.~\ref{figpi1} is
displayed in a similar manner as Fig.~\ref{fig1},
and the former may be formally regarded as representing the case for a ``nucleon'' assumed to possess the fictitiously large sigma terms
such that $\sigma_{\pi N}+ \sigma_{s} \to M/2$, corresponding  to (\ref{replacenpi}),
which results in the considerable increase of the LO value due to the contribution from the second term of (\ref{ccc}).
The asymptotic value of (\ref{ccc}) now becomes
\begin{equation}
-0.145556+0.305556 \frac{\langle \pi(p)|m\bar{\psi}\psi|\pi(p)\rangle}{2 m_\pi^2}
\simeq 0.007\ , 
\label{asymptoticvpi}
\end{equation}
and this small value due to the cancellation leads to the small values at LO in Fig.~\ref{figpi1}(a),
to which the NLO and NNLO corrections give ten \%-level and \%-level effects, respectively.
Fig.~\ref{figpi1}(b) shows that the anomaly terms are now positive and much larger than the corresponding contribution in
Fig.~\ref{fig1}(b); this contribution strongly cancel the negative twist-2 effect, resulting in the rather small total value.
Thus, Fig.~\ref{figpi1} shows a quite different pattern, compared to Figs.~\ref{fig1}, \ref{fig2} for the nucleon case.

\begin{figure}[hbtp]
\begin{center}
\includegraphics[width=0.5\textwidth]{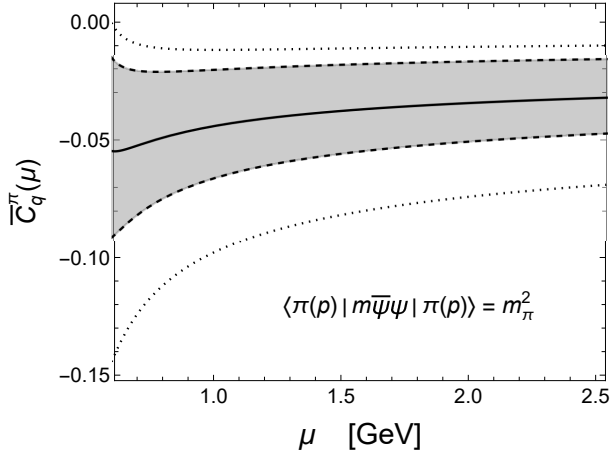}
\caption{The 
pion's gravitational form factor $\left. {\bar C}^\pi_q(\mu )\right|_{n_f=3}$ at the NNLO (3-loop) accuracy,
given by (\ref{ccc}) with (\ref{replacenpi}):
the shaded areas indicate
the uncertainties estimated by varying $\frac{1}{4}{\Theta}_{2q}\left(\mu _0=1.3~{\rm GeV}\right)$ within 
the uncertainties of $\frac{1}{4}{\Theta}_{2q}\left(\mu _0=1.3~{\rm GeV}\right)= 0.61\pm 0.08$;
the solid curve is same as the solid curve in Fig.~\ref{figpi1}(a) (i.e., the NNLO result using $\frac{1}{4}{\Theta}_{2q}\left(\mu _0=1.3~{\rm GeV}\right)=0.613$); also shown by the upper and lower dotted lines are
the NNLO results using $\frac{1}{4}{\Theta}_{2q}\left(\mu _0=1.3~{\rm GeV}\right)=0.5$ and
$0.8$, respectively.
}
\label{figpi2}
\end{center}
\end{figure}

From the results in Fig.~\ref{figpi1},
we expect that the uncertainties in our calculation of $\left. \bar{C}^\pi_q(\mu)\right|_{{n_f} = 3}$
due to the omission of the terms of N$^3$LO and higher should be $\lesssim\ $ a few~\%.
It is also known that the corrections to (\ref{chiralp}), (\ref{chiralpp}) by
chiral perturbation theory is very small ($\lesssim 6$~\%)~\cite{Gasser:1983yg,Colangelo:2001sp,Scherer:2012xha};
therefore,
the uncertainty in the present calculation of $\left. \bar{C}^\pi_q(\mu)\right|_{{n_f} = 3}$ appears to be dominated by
the uncertainties exhibited in (\ref{theta2qinput}).
In view of this, the shaded area in Fig.~\ref{figpi2} displays 
the uncertainties estimated by varying $\frac{1}{4}{\Theta}_{2q}\left(\mu _0=1.3~{\rm GeV}\right)$ within 
the uncertainties of $\frac{1}{4}{\Theta}_{2q}\left(\mu _0=1.3~{\rm GeV}\right)= 0.61\pm 0.08$, and 
the solid line is same as the solid line in Fig.~\ref{figpi1}(a);
some explicit values with the uncertainties corresponding to the shaded area read 
\begin{eqnarray}
\left. {\bar C^\pi_q(\mu=0.7~{\rm GeV} )} \right|_{{n_f} = 3}&=&-0.05\pm 0.03\ ,
\nonumber\\
\left. {\bar C^\pi_q(\mu=1~{\rm GeV} )} \right|_{{n_f} = 3}&=&-0.04\pm 0.02\ ,
\nonumber\\
\left. {\bar C^\pi_q(\mu=2~{\rm GeV} )} \right|_{{n_f} = 3}&=&-0.03\pm 0.02\ .
\label{explicitresultspi}
\end{eqnarray}
These are the values in the $\overline{\rm MS}$ scheme.
These results indicate that the behaviors of $\bar{C}^\pi_q(\mu)$ for the pion are quite different from those of $\bar{C}_q(\mu)$
for the nucleon, and, in particular, the absolute magnitude of the former is much smaller than that of the latter, see (\ref{explicitresults}).

It is remarkable that the Nambu-Goldstone nature of the pion allows us to determine the matrix element of the
quark scalar operator accurately, as (\ref{chiralpp}), although the corresponding quantities for the nucleon case, the sigma terms, are the major source of the uncertainty to calculate the nucleon's $\bar{C}_q(\mu)$.
The uncertainties in Fig.~\ref{figpi2} and (\ref{explicitresultspi}) reflect those of the input for $\frac{1}{4}{\Theta}_{2q}\left(\mu _0\right)$,
(\ref{theta2qinput}).
Our NNLO formula, (\ref{ccc}) with (\ref{replacenpi}), would allow us to predict the value of 
the pion's $\left. \bar{C}^\pi_q(\mu)\right|_{{n_f} = 3}$
at the accuracy of $\sim$ percent level, when the value of $\frac{1}{4}{\Theta}_{2q}\left(\mu _0\right)$ 
were fixed at the NNLO level by global QCD analysis or by lattice QCD.

\section{Conclusions}
\label{conc}

In this paper we have presented the NNLO QCD calculation of the forward value of the twist-four gravitational form factor $\bar{C}_{q,g}$.
Our model-independent calculation is based on 
exact QCD constraints on $\bar{C}_{q,g}$, provided by an extended version of the QCD trace anomaly,
such that the trace anomaly is attributed to the anomalies arising in each of the quark part and gluon part of the QCD energy-momentum tensor.
This allows us to reexpress the forward value of $\bar{C}_q$ for the nucleon 
in terms of the target mass effect associated with the average value of the quark momentum fraction, 
and in terms of the expectation value of the quark contribution 
of the trace anomaly. The forward value of $\bar{C}_g$ can be expressed similarly using the corresponding gluonic quantities,
and the fact that the QCD trace anomaly equals the sum of the quark anomaly and gluon anomaly ensures the relation, $\bar{C}_q+\bar{C}_g=0$.
Using the three-loop DGLAP evolution of the quark momentum fraction and the three-loop formula of the quark/gluon decomposition of the QCD trace anomaly, we derive the NNLO formula of the forward value, $\bar{C}_q(\mu)$, which exhibits the dependence on
the renormalization scale $\mu$.
This NNLO formula coincides with the solution of the three-loop RG equation for the twist-four quark-antiquark-gluon operator
whose matrix element gives $\bar{C}_q(\mu)$. The terms of this formula are organized clarifying each order of the LO, NLO, and NNLO 
in the RG-improved perturbation theory; for this purpose,
taking into account the nucleon mass formula derived from the QCD trace anomaly plays essential roles, so that
the matrix elements of the operator $F^2$ arising from the quark contribution to the trace anomaly are reexpressed in favor of the 
nucleon mass and the matrix elements of the quark scalar operator $m\bar{\psi}\psi$.

As a result, our NNLO formula for $\bar{C}_q(\mu)$ involves, apart from the nucleon mass,  the two types of nonperturbative parameters: 
the quark momentum fraction $A_q$ corresponding to twist-two effect and the sigma terms corresponfing to the twist-four operator $m\bar{\psi}\psi$.
As a remarkable point of the formula, it has the $\mu$-independent constant terms 
that are determined completely by $N_c$ and $n_f$.
Those constant terms represent the asymptotic value of $\bar{C}_q(\mu)$ as $\mu \to \infty$ in the chiral limit,
and are composed of the contribution due to 
the asymptotic value of the quark momentum fraction $A_q$ and of the contribution originating from the behavior 
$\left\langle N(p) \right| F^2   \left| N(p)  \right\rangle\sim M^2/\alpha_s$ in the quark anomaly effect.
Although the approach of $\bar{C}_q(\mu)$ to the corresponding asymptotic value is quite slow due to the RG evolution effect of $A_q$, 
this asymptotic value determines the model-independent ``basis value'' for the NNLO estimation of $\bar{C}_q(\mu)$.
We find that the nonperturbative parameters participate in our NNLO formula accompanying the small numerical coefficients,
so that the nonperturbative parameter $A_q$ as well as the sigma terms produces at most 30~\% level modification.
As the result, 
the NLO as well as the NNLO perturbative corrections associated with $A_q$ yield the percent-level corrections to the LO evaluation,
and the uncertainties in the input values of the sigma terms lead to only a few percent uncertainties in our evaluation of 
$\bar{C}_q(\mu)$, allowing us to obtain accurate NNLO prediction in the $\overline{\rm MS}$ scheme, 
$\left. {\bar C_q(\mu=1~{\rm GeV} )} \right|_{{n_f} = 3}=-0.180\pm 0.003$.
We find that the $\mu$ dependence is significant in the relevant region, $0.7~{\rm GeV} \lesssim \mu \lesssim 2~{\rm GeV}$,
for which we provide a simple approximate formula to reproduce the $\mu$ dependence at NNLO.

We also extend those results 
to the case of the spin-0 hadrons, in particular, a pion.
In the context of evaluating our NNLO formula, the pion may be formally regarded as a ``nucleon'' assumed to possess fictitiously large sigma terms, whose value are determined precisely by the Nambu-Goldstone nature of the pion.
The corresponding large sigma terms lead to cosiderable positive modification to the asymptotic ``basis value'',
so that our NNLO evaluation indicates the nonzero but small value,
$\left. {\bar C^\pi_q(\mu=1~{\rm GeV} )} \right|_{{n_f} = 3}=-0.04\pm 0.02$, in the $\overline{\rm MS}$ scheme.
The significant uncertainty of this prediction reflects the uncertainties in the average value of the quark momentum fraction 
in the pion based on the recent NLO global fits of the pion's parton distribution functions.
Those results, compared with those for the nucleon, 
indicate quite different pattern, revealed as a new aspect by exploiting 
the quark/gluon decomposition of the QCD trace anomaly.

The present result may have implications on 
the spin sum rule for the nucleon~\cite{Ji:1996ek}, in particular, for the transversely polarized case:
the quark/gluon total angular momentum $J_{q,g}$ are expressed as
\beq
J_{q,g}=\frac{1}{2}(A_{q,g}+B_{q,g}) + f(p_z)\bar{C}_{q,g}
\eeq
where $f(p_z)=0$ for the longitudinally polarized nucleon, while, for the transversely polarized nucleon,   
$f(p_z)$ is a frame-dependent function (depends on the nucleon longitudinal momentum $p_z$) which vanishes at $p_z=0$ and approaches $\frac{1}{2}$ as $p_z\to \infty$~\cite{Hatta:2012jm,Leader:2012ar,Chakrabarti:2015lba} (see also \cite{Ji:2012vj}). 
It was noted~\cite{Hatta:2018sqd} that, asymptotically, $\frac{1}{2}(A_q+B_q) \approx 0.18$, while $\bar{C}_q=-\bar{C}_g \approx -0.15$ for $n_f=3$,
indicating the effect of the last term could be significant.
Now it is confirmed and improved by the present result, 
$\left. {\bar C_q(\mu=1~{\rm GeV} )} \right|_{{n_f} = 3}=\left.- {\bar C_g(\mu=1~{\rm GeV} )} \right|_{{n_f} = 3}=-0.180\pm 0.003$.

Our result could be useful also for the studies of the quark/gluon contributions of pressure distributions inside the hadrons,
the near-threshold photoproduction of $J/\psi$ in $ep$ scattering, and the origin of the hadron mass.
Our NNLO prediction may be compared with the future direct calculations of $\bar{C}_q(\mu)$ in lattice QCD.
Also, the present result should impose the constraints on the studies of the $t$ dependence of the gravitational form factor
$\bar{C}_{q,g}(t, \mu)$, providing its normalization at $t=0$.

\section*{Acknowledgments}
The author thanks Shunzo Kumano, Hiroyuki Kawamura, and Yuichiro Kiyo for insightful discussions.
This work was supported by JSPS KAKENHI Grant Number JP19K03830.

\bibliography{../../../Apps/Overleaf/pion/pion}

\end{document}